\documentclass[twocolumn,aps,preprintnumbers,letterpaper]{revtex4}
\usepackage{amsmath,amssymb}
\usepackage{epsfig}
\usepackage{graphicx}
\usepackage{amsmath}
\usepackage{slashed}
\usepackage{amsfonts}
\usepackage{epstopdf}
\usepackage{color}
\usepackage[caption=false]{subfig}
\usepackage[pdftex, pdfborder= 0 0 0, citecolor=blue, urlcolor=blue, linkcolor=blue, colorlinks=true, bookmarksopen=true]{hyperref}
\newcommand{\be}{\begin{equation}}
\newcommand{\ee}{\end{equation}}
\newcommand{\bea}{\begin{eqnarray}}
\newcommand{\eea}{\end{eqnarray}}
\newcommand{\ba}{\begin{eqnarray}}
\newcommand{\ea}{\end{eqnarray}}

\newcommand{\Dslash}{D\hspace{-1.6ex}/\hspace{0.6ex} }

\newcommand{\pslash}{p\hspace{-1.6ex}/\hspace{0.6ex} }
\newcommand{\partslash}{\partial\hspace{-1.6ex}/\hspace{0.6ex} }

\begin{document}

\title{ Sphalerons,  baryogenesis and helical magnetogenesis \\
in the electroweak transition of the minimal standard model}
\author{ 
 Dmitri Kharzeev$^{1,2}$, Edward Shuryak$^1$ and Ismail Zahed$^1$ }
 
\affiliation{$^1$ Department of Physics and Astronomy, \\ Stony Brook University,\\
Stony Brook, NY 11794, USA}
\affiliation{$^2$ Physics Department and RIKEN-BNL Research Center, \\
Brookhaven National Laboratory, \\
Upton, NY 11973, USA} 

\begin{abstract} 
We start by considering the production rates of sphalerons with different size $\rho$ in the symmetric phase, $T>T_{EW}$. At small $\rho$, 
the distribution is cut off by the growing mass $M\sim 1/\rho$, and at large $\rho$ by the magnetic screening mass. 
In the broken phase, $T<T_{EW}$ the scale is set by the  Higgs VEV $v(T)$.  We introduce the concept of 
"Sphaleron freezeout" whereby the sphaleron production rate matches 
the Hubble Universe expansion rate. At freezeout the sphalerons are out of equilibrium. Sphaleron explosions
 generate sound and even gravity waves, when nonzero Weinberg angle make them non-spherical.  
 We revisit  CP violation during the sphaleron explosions. We assess its magnitude using the Standard Model CKM quark matrix,
 first for nonzero and then zero Dirac eigenstates.  We find that  its magnitude is maximal
at the sphaleron freezeout condition with $T\approx 130\, GeV$. We proceed to estimate the amount of 
CP violation needed to generate the observed magnitude of baryon asymmetry of Universe. 
The result is about an order of magnitude below our CKM-based estimates. 
We also relate the baryon asymmetry to the generation of  $U(1)$  magnetic chirality, 
which is expected to be conserved and perhaps visible  in polarized intergalactic magnetic fields.
 \end{abstract}

\maketitle

\section{Introduction}
One of the central unsolved problems of cosmology  is baryogenesis,  the
explanation of the apparent baryon asymmetry (BAU)  in our Universe.
Since the problem involves many areas of physics, our introduction will be split into several parts for
their brief presentation. The common setting is the cosmological Electroweak Phase Transition (EWPT), 
whereby the universe undergoes a transition from a symmetric phase to a broken phase  with a nonzero 
vacuum expectation value (VEV) for the Higgs field.

Most of the  studies of the EWPT, from early works till now,  assumed the phase transition 
to be first order, producing bubbles
with large-scale deviations from equilibrium~\cite{Witten:1984rs}. Most studies of gravitational radiation 
were carried in this setting.   However, Ref.\cite{Kajantie:1996mn} and subsequent lattice studies have shown that the standard model (SM) 
can only undergo a first order transition for  Higgs masses well below the $M_H\approx 125$ GeV mass observed at LHC. 
The first order transition remains possible only in the models that go   beyond the standard model   (BSM), which we do not discuss in this work.

Another possible scenario of the EWPT  is the  ``hybrid" or ``cold" scenario, suggesting that the broken 
symmetry phase happens at the end of the inflation epoch.  Here, the  label  ``cold"  refers to  the fact that 
at the end of the reheating and equilibration of the Universe, the temperature becomes of the order of  $T=30-40$ GeV, 
well  below the critical electroweak temperature $T_{EW}\approx 160$ GeV.
Violent deviations from equilibrium occure
 in this  scenario~\cite{GarciaBellido:1999sv,Krauss:1999ng}.
Detailed numerical studies~\cite{GarciaBellido:1999sv,Smit_etal} revealed ``hot spots", filled with strong  gauge field, 
 later identified \cite{Crichigno:2010ky} with certain multi-quanta bags
containing gauge quanta and top quarks. We do not consider this scenario in this work as well.

Instead,  we will focus on the least violent scenario for the EWPT,
a smooth  crossover transition expected from the Minimal Standard Model.  
The main cosmological parameters of the EWPT are by now  well established. For completeness
and consistency, they are briefly summarized in  Appendix A.

\subsection{Sphalerons near the EWPT} \label{intro_sphalerons}

Sphaleron transitions are topologically nontrivial fluctuations of the gauge field with changing  Chern-Simons number
$N_{\rm CS}$.  The multidimensional effective potential 
$V(N_{\rm CS})$ possesses a one-parameter sphaleron path,  
along which thermal fluctuation can cause
 a slow ``climb" uphill, to the saddle points at half-integer $N_{\rm CS}$, the $sphaleron$
 (e.g.  from $N_{CS}=0$ to $N_{CS}=\frac 12$). The explicit static and purely magnetic gauge 
 configuration was originally found in~\cite{Klinkhamer:1984di}. When perturbed, 
 the saddle point configuration leads to a classical roll down in the form of a  time-dependent
solution known as the  ``sphaleron explosion". 
 The Chern-Simons number in such processes changes back from half-integer to integer (e.g.
from $N_{\rm CS}=\frac 12$ to $N_{\rm CS}=1$ or 0).

In the symmetric (unbroken) phase with $v(T>T_{EW})=0$, 
the sphaleron rate is only suppressed by the (5-th) power of the coupling \cite{Kuzmin:1985mm,Arnold:1996dy}
, without exponent 
$\Gamma/T^4\sim \alpha_{EW}^5\sim 10^{-7} $.
In the broken phase after the EWPT with $v(T<T_{EW})\neq 0$, it is suppressed further,
by a Boltzmann factor ${\rm exp}[-M_{sph}(T)/T]$ with a T-dependent sphaleron mass.
 Some basic information about the electroweak sphaleron rate
is given in  Appendix B. We start this work  by
discussing the sphaleron size distribution. This is done separately for (a) unbroken and (b) broken phases
(with nonzero Higgs VEV $v(T)\neq 0$). 

In case (a) one can ignore the Higgs part of the action and focus on the gauge part.
At small sizes the distribution is cut off because the sphaleron mass is increasing $m\sim 1/\rho$ (by dimension), with the pure gauge 
solution  discussed in~\cite{Ostrovsky:2002cg} and~\cite{Shuryak:2002qz}. At large sizes, 
the limiting factor is the magnetic screening mass which we will extract from  lattice data~\cite{Heller:1997vha}. We then interpolate
between the ``large size" and ``small size" expressions, with the overall normalization
of the rate tuned to available lattice data \cite{DOnofrio:2014rug}.

In case (b) we follow the original  work in~\cite{Klinkhamer:1984di}, by inserting  the
appropriate parameters of the effective electroweak action neat $T_{EW}$ calculated in~\cite{Kajantie:1995dw}.
Specifically, we use a one-parameter Ansatz depending on the parameter $R$, for which both the mass $M(R)$ and
the r.m.s. size $\rho(R)$ are calculated. The results will be summarized in Fig.\ref{fig_P_of_rho} below.

\subsection{Generation of sounds and gravity waves} \label{intro_sounds_gravity}

The ``sphaleron explosion" is described by a time-dependent solution
of the classical Yang-Mills equations. A number of such solutions have  been obtained numerically.
Analytic solutions  for pure-gauge sphalerons have  been obtained in~\cite{Ostrovsky:2002cg} and  in~\cite{Shuryak:2002qz},
 of which we will use the latter one. Some details of how it was obtained and some
basic formulae are summarized in Appendix~\ref{appendix_pure_gauge_sphalerons}.

As we will see below, the word ``explosion" is not a metaphor here. Indeed, 
the time evolution of the stress tensor $T^{\mu\nu}(t,\vec x)$ does display  an expanding shell of energy. 
Although we have not studied its interaction with  ambient  matter in any detail, it is clear that a significant 
fraction of the sphaleron mass should end up in spherical sound waves. 

In the symmetric phase, the sphaleron explosion is spherically symmetric. 
It does not sustain a quadrupole deformation and therefore cannot radiate direct gravitational 
waves. However, the indirect gravitational waves can still be generated at this stage, 
through the process$$ {\rm sound+sound \rightarrow gravity \,\,wave}$$
  pointed out in~\cite{Kalaydzhyan:2014wca}. After the EWPT, at
 $T<T_{EW}$, the nonzero VEV breaks the symmetry and the sphalerons (and their explosions) are no longer spherically symmetric. 
 With a nonzero and time-dependent quadrupole moment, they generate
 direct  gravitational radiation. We will calculate the corresponding matrix elements 
 of the stress tensor in section \ref{gravity_waves}.

\subsection{Baryon asymmetry of the Universe (BAU)} \label{intro_BAU}
 Any explanation of the baryon asymmetry in the universe (BAU) needs, as noted by
Sakharov  long ago \cite{Sakharov:1967dj},   three key conditions:  1/ deviation from equilibrium; 2/ baryon number violation; 3/ CP violation. 

It is well known that the standard model (SM) includes all these conditions ``in principle".
In particular, since the baryon number is locked to the Chern-Simons number, sphaleron explosions
produce $\pm 3$ units of baryon and lepton numbers. The CP violation does happen, due to
complex phase of the CKM quark matrix. 
 However, specific scenarios based on SM were so far unable to reproduce the key
observed  BAU  parameter, the baryon-to-photon ratio 
\be 
\frac {n_B}{n_\gamma}\sim 6\cdot 10^{-10}  \label{eqn_ba}
\ee
at the time of primordial nuclearsynthesis. 

 As a result, the mainstream of BAU studies has shifted mostly to scenarios containing
 unknown physics  ``beyond the Standard Model" (BSM), in which  hypothetical sources of CP violation are introduced (e.g. axion fields, or extended Higgs or neutrino sector
with large CP violation.)  The so called $leptogenesis$ scenarios use superheavy neutrino decays, occurring at very high scales,
and satisfying both large CP and out-of-equilibrium requirements, with large lepton asymmetry
 transformed into the baryon asymmetry by the electroweak sphalerons at $T_{EW}$.  
While one of these BSM scenarios may well turn out to be the explanation for BAU, 
they still remain purely hypothetical at this time, lacking any support  from current  experiments. 

 The aim of this work is to provide a scenario {\em within the SM}  that maximizes 
 BAU. We will estimate,  as accurately as  possible at this time,  the 
magnitude of BAU that  the SM predicts. 
Throughout, we will keep to a conservative and minimal SM (MSM) scenario, in which the EWPT is smooth, with gradual
building of the Higgs VEV $v(T)$ at $T<T_c$. The needed ``out-of-equilibrium" conditions 
to be discussed below, will be associated with ``sphaleron freezeout" of large-size sphalerons, with the rates 
 comparable  with the universe expansion rate. Contrary to popular opinion, it turns out to be only an order of magnitude below  what is phenomenologically needed. We therefore think that this scenario deserves much more detailed and quantitative studies.

 

\subsection{Introductory discussion of CP violation in the Standard Model } 
\label{intro_CP}

The CP violation in the SM is induced by the nonzero phase of the 
Cabbibo-Kobayashi-Maskawa (CKM) matrix. Its magnitude is known to be strongly scale dependent.
Naively, at $T_{EW}$ all particle momenta are of the order of $p\sim 3T\sim 500$ GeV, above all quark masses.
As originally shown by Jarlskog~\cite{Jarlskog:1985ht},  at such high scale the magnitude of the CP violation needs to be proportional 
to the product of two different factors.
The first is the so called ``Jarlskog determinant" $J$ containing the sine of the CP violating phase
 $sin(\delta)$,  times sines and cosines
 of the mixing angles.  $J$ has a geometric meaning, so
 it is invariant under re-parameterization of the CKM matrix. 
 Its numerical value is $J\sim 3\cdot  10^{-5}$.
The second factor is the (dimension 12) product of the string of squared up and down quark mass differences

\be 
(m_b^2 - m_d^2) (m_b^2 - m_s^2) (m_d^2 - m_s^2) \label{eqn_mass_differences}
\ee
$$ \times(m_c^2 - m_t^2) (m_c^2 - m_u^2) (m_t^2 - m_u^2) $$

which ensures  that  CP asymmetry vanishes whenever any two masses  of up-kind or down-kind quarks are equal. 
The resulting CP asymmetry at electroweak momentum scale is  very small, $A_{CP}\sim 10^{-21}$~\cite{Shaposhnikov:1987tw} ,  
dashing naive expectations for the SM to significantly contribute to BAU. 
Our calculation of the effective CP-violating Lagrangian at the beginning of section~\ref{CP} 
agrees with Jarlskog argument just presented above, and specifically Shaposhnikov's estimate~\cite{Shaposhnikov:1987tw}. 

And yet, this is not the last word on the issue: people look for ways to go around the Jarlskog argument, with this paper being one of them. (A somewhat analogous situation in physics arose when people naively used direct reaction rates
for burning of hydrogen into helium in stars. The cross sections were smaller than
needed, by many orders of magnitude. Only with time and work, Bethe's nontrivial 
chains of reactions were eventually found and  explained why the Sun shines.).  

    In fact the string of mass differences (\ref{eqn_mass_differences})
   divided by the 12-th power of the momentum scale $p \sim \pi T$,  enters only when $p$ is above all quark masses. If the relevant momentum scale is different, CP asymmetry can be much larger. 
It is known (and
 we will show it in section \ref{CP} as well), that the asymmetry turns out to be maximal at
the momentum scale $p\sim 1$ GeV, in a ``sweet spot"  between the masses of the light and heavy quarks,  with  15 orders of CP suppression gone!
Scenarios using quarks at  a momentum scale $p\sim 1$ GeV were suggested in~\cite{Farrar:1993hn},
but then criticized and found untenable, as a small momentum is not possible  to keep for a long time. 

   More importantly, one may question whether  the product of the string of 
   mass differences (\ref{eqn_mass_differences})  found in the evaluation of the effective Lagrangian needs to
   be universally present in any CP violating process. Clearly, this cannot be the case for
otherwise we would never be able to observe it experimentally. 
   
   The  CP violation 
 originally discovered in neutral $K$ decays is of magnitude
$\sim 10^{-3}$, but these processes are complicated by relation to $K^0-\bar K^0$ mass difference. 
Consider the much simpler case of ``direct CP violation" in exclusive charged mesons (or baryon) decays,
induced by reactions of the type $b\rightarrow \bar q q q'$ (e.g. $b\rightarrow \bar c c s$) or many others.
The tree diagram of the decay has two CKM matrix elements, and CP violation comes from the interference with the
so called ``penguin" diagrams, with an additional gluon, producing $ \bar q q$ pair. 
This second diagram has in general a sum over up-type quarks $U$ and therefore 
CKM matrix elements $V_{bU}$ and $V_{Uq'}$. CP asymmetry is not only observed, but
is rather large,  suppressed only by the
strong coupling constant and some numerical smallness of a loop, like $1/4\pi$.  

Our main point here is the following:  out of the three 
generation of down quarks, only two (say $b,s$) are involved, while the remaining one
(say $d$) is $not$. This means, whatever the expression for CP asymmetry may be, it cannot
possible contain $m_d$. Therefore,  factors such as $ (m_b^2 - m_d^2)  (m_s^2 - m_d^2)$ expected from
theoretical argument given above cannot be there,  so this argument is not really universal. 

And yet, we do know that in Universe with $m_d=m_b$ or $m_d=m_s$ there should be no CP violation! 
How could this be? The point we want to make from this example is
 that  the quark masses and the CKM matrix elements  are {\em not 
independent} of each other.  The CKM matrix $somehow$ knows by itself
what features to have, in a Universes with degenerate quarks. 

In some models one can see how this relation takes place. Instead of discussing them, we 
emphasize    that the {\em experimentally determined} entries in the CKM matrix in
{\em our Universe} are such that they produce CP violation in $b$ decays.  
The nontrivial CP violating relative phase between two pairs of CKM matrix elements tells us that
the $d$ quark is $not$ degenerate with others. We can learn this 
even from processes in which no $d$ quarks
participate!  ( While we will not  discuss CP violation in quark decays below,
this lesson  will be relevant to our results about sphaleron production of $d$ and $u$
quarks.)   

Let us now briefly describe 
our approach to CP violation. It consists of two parts, different in physics and method, both in section \ref{CP}. 
The first starts with the traditionally set problem of an effective one-quark-loop
Lagrangian in a background of an arbitrary gauge field.
Our approach has some differences with those of others, in that we
 use a basis of  Dirac eigenstates instead of momenta. It  leads certain universal functions of
 quark masses and eigenvalues $\lambda$, to be convoluted with a background-specific
 spectral density $\rho(\lambda)$. 
 We reproduce several known facts about such Lagrangian, and also discuss
inclusions  of background $Z$ and electromagnetic fields. In the case of CP violation those
are not just corrections: instead those inclusions are crucial to break certain cancellation
patterns and to produce a nonzero CP violation.
 
 Our plan  is to use such approach in the background field corresponding to the sphaleron explosion.  
The eigenvalue spectrum is different from the momentum spectrum, as it contains both nonzero and zero modes, the latter known to be of topological origin. 
The background of a sphaleron explosion is analytically known only at zero temperature and
without the  Higgs VEV,  but we suggest that the {\em topological stability} condition
preserves the  Dirac zero mode (and thus baryon number violation) even in the
presence of a finite-$T$ plasma perturbations (like quark scattering off thermal gluons).
 The magnitude  of CP violation corresponding to  a generic zero Dirac eigenvalue follows 
from the (flavor-dependent) phases of  the outgoing quark waves (the zero mode itself).
We show that the CP violation in the exclusive production of $u$ and $d$ quarks is of magnitude $10^{-9}$,
vanishing in the sum, but still producing an  asymmetry of this magnitude due to presence of 
electromagnetic fields.

\subsection{ Intergalactic magnetic fields and helical magnetogenesis from EWPT}
\label{intro_helical}

The EWPT has also been suggested to be a source for large scale magnetic fields
in the universe. The existence and properties of  intergalactic magnetic fields are hotly 
debated by observational astronomers, cosmologists and experimentalists specialized in the
detection of  very high energy cosmic rays. Currently, there are lower and upper limits on the magnitude of these fields.
The issues of the chirality of these fields as well as their correlation scale are still open questions, with suggestions
ranging from  larger than the visible size of the universe  (in case of pre-inflation chiral fluctuations) to sub-Galaxy size. 
Many things {\em may} happen on the way from the Big Bang to today$^\prime$s magnetic fields.

Our main point is that the sphaleron-induced BAU must also be related with the chiral imbalance of quarks and leptons
 produced in sphaleron transitions. This chiral imbalance is then transferred to linkage of magnetic
 fields. Since the linkage is expected to be conserved in plasmas, it may be observable  today, via preference of certain magnetic helicity in the intergalactic magnetic fields.

%
%




\section{Sphalerons near the crossover EW phase transition }

\subsection{The temperature dependence of the sphaleron rates}  \label{rates}

To assess the temperature of the sphaleron rate, we first start in the symmetric phase with zero Higgs VEV and $T>T_{EW}$.
The change in the baryon number is related to the sphaleron rate through~\cite{Khlebnikov:1988sr,Moore:2000ara},

\be
 {1 \over N_B}{dN_B \over dt}={39 \Gamma \over 4T^3} 
  \label{eqn_Moore}  
 \ee
The sphaleron rate calculated from earlier  lattice studies and also derived from Bodeker model  is

\be
 \Gamma=\kappa\, \bigg( { gT \over m_D} \bigg)^2 \alpha_{W}^5 T^4
 \ee
with $\kappa\sim 50$ extracted from the lattice fit. 
The  lattice work~\cite{DOnofrio:2014rug}  yields a more accurate evaluation for the rate

\be 
\label{NUMBER}
{\Gamma \over T^4}=(18\pm 4)\alpha_{EW}^5 \approx 1.5\cdot 10^{-7}
 \ee
While (\ref{NUMBER})  appears small, its folding in time at the electroweak transition temperature
$T_{EW}$ is large

\be   
{1 \over N_B}{dN_B \over dt} t_{EW}=3.2 \cdot 10^{9} 
\label{eqn_rate_unsuppressed}
\ee
Therefore, the  baryon production rate in the symmetric phase  strongly
exceeds the expansion rate of the Universe $H\sim 1/t_{EW}$,  by 9 orders of magnitude!
 Therefore, prior to EWPT,
  $T\geq T_{EW}$, the sphaleron transitions are in thermal equilibrium.
According to Sakharov, this excludes the  formation of BAU. In fact, this even suggests
a total wash of baryon-lepton (BL) asymmetry. This particular conclusion will be circumvented below, by ``sphaleron freezeout" phenomenon.

Another important result of  the lattice work~\cite{DOnofrio:2014rug} is the temperature dependence of the sphaleron rate
in the symmetry broken phase

\begin{widetext}
\begin{eqnarray}
\label{lattice_rate} 
{\rm Log}\bigg({\Gamma(T<T_{EW}) \over T^4}\bigg)= 
 -(147.7\pm 1.9) +(0.83\pm 0.01) \bigg({T \over {\rm GeV}}\bigg) \nonumber\\
 \end{eqnarray}
 \end{widetext}
%
%
%

It would be useful for our subsequent discussion,  to re-parametrize  this rate, expression it in terms of the sphaleron mass 
through  the temperature-dependent Higgs VEV  $v(T)$, namely 

\be 
\label{DMV}
\frac{\Gamma}{T^4}\sim {\rm exp}\bigg(- \frac{\Delta M_{v}}T\bigg)
\ee 
with
\be 
\Delta M_{v}(T)\approx  {v(T)^2 \over 9\, {\rm GeV}} \label{delta_M_v} 
\ee

By comparing this rate to the Hubble value for the Universe expansion rate at the time $t_{EW}$,   
\cite{DOnofrio:2014rug} concluded that the sphaleron transitions  become irrelevant  when
temperature is below
\be 
T_{\rm decoupling}=131.7 \pm 2.3 \, {\rm GeV}
 \ee
So, our subsequent discussion is limited to the times when the temperature is 
in the range 
$$ T_{EW} \approx 160\, {\rm GeV} < T< T_{\rm decoupling}\approx 130\, {\rm GeV} $$
Note that by this time, the Higgs  VEV (\ref{lattice_v_of_T}) reaches only a fraction 
of its value today, in the fully broken phase, i.e. $v(T=0)\approx 246\, {\rm GeV}$.

\subsection{The sphaleron size distribution} 
\label{size_distribution}

The lattice results recalled above, gave us valuable information of the 
mean sphaleron rates and thus masses. However for the purposes of this work,
we need to know  their size distribution.
As we will detail below, baryogenesis driven by CP violation is biased toward sphalerons 
of sizes larger then average, while gravity wave signal and seeds of magnetic clouds are
biased to smaller sizes.

\subsubsection{Unbroken phase and small sizes}

Let us start with the small-size $\rho$ part of the distribution. In this regime, we can ignore the
Higgs VEV,  even when it is non-vanishing, a significan simplification.
By dimensional argument it is clear that $M_{\rm sph}( \rho) \sim 1/\rho$.
It is also clear that small-size sphalerons should be spherically symmetric. 

The classical sphaleron-path configurations 
in pure gauge theory were analytically found in \cite{Ostrovsky:2002cg}.  The method used is ``constrained minimization" of
the energy, keeping their size  $\rho$ and their Chern-Simons number $N_{CS}$ fixed. 
This gave the explicit shape of the sphaleron barrier. At the highest point of the barrier 
 $N_{CS}=\frac 12$, the sphaleron mass is
 
\be 
M_{\rm sph}( \rho)= {3\pi^2 \over g^2 \rho}  
\label{small_rho} 
\ee
Later the same solutions were obtained  in ~\cite{Shuryak:2002qz}
by a different method,
 via an off-center conformal transformation of the Euclidean solution
(the instanton)  of the Yang-Mills equation. Some of the results are reviewed in  Appendix B. 
It provides not only a static sphaleron configuration, but the whole {\em  sphaleron explosion
process} in relatively simple analytic form, to be used below. 


\subsubsection{Unbroken phase and large sizes}


Now we turn to the opposite limit of large-size sphalerons. 
Since the sphaleron itself is a magnetic configuration, at large $\rho$
one should consider magnetic screening effects.
Unlike the simpler electric screening, the magnetic screening does not appear
in perturbation theory~\cite{Polyakov:1978vu}.
It is purely nonperturbative, and likely due to  magnetic monopoles.

The magnetic mass  $M_m$ conjectured by 
 Polyakov to scale as $M_m={\cal O}( g^2 T )$, was confirmed by lattice studies.
While in the QCD plasma the coupling is large and the difference between
the electric and magnetic masses is only a factor of two or so, in the electroweak plasma the coupling is small
$\alpha_{EW}\sim 1/30$,   and therefore the magnetic screening mass is 
 smaller than the thermal momenta by about two order of magnitude

\be
{M_m \over 3T}\sim {\alpha_{EW}  \over 3} \sim 10^{-2} 
\ee
The key consequence for the sphalerons is that their sizes would be about two orders of magnitude larger than
the interparticle distances in the electroweak plasma. This conclusion, in turn, will have dramatic consequences 
for the magnitude of the CP violation.

The part of the gauge action related with
the screening mass is 

\begin{eqnarray}
\label{SS}
\Delta S_{\rm screening}={M^2_m\over 2} \int d^4x  (A_i^a)^2 
\end{eqnarray}
For static sphalerons, the 
integral over the Matsubara time is trivial,  giving $1/T$. Parametrically,
we have  $M_m\sim g^2 T, A\sim 1/g \rho$,  so that

\begin{eqnarray}
&&M^2_m\, \int d^4x  (A_i^a)^2\sim \nonumber\\
&&(g^2 T)^2 \bigg({1 \over g \rho}\bigg)^2 {\rho^3 \over T} \sim g^2 T \rho 
\end{eqnarray}
At high temperature, the pure SU(2) lattice simulations  in~\cite{Heller:1997vha}  give

 \be 
 \label{MLATTICE}
 M_m(T) \approx  0.457 g^2 T
 \ee
Inserting (\ref{MLATTICE}) in (\ref{SS})  and using the pure gauge sphaleron configuration
yield the ecreening factor for large size sphalerons

  \be 
  {\Gamma \over T^4}\sim {\rm exp}\bigg(- (0.457)^2\pi^2 g^2 T \rho\bigg)  
  \label{large_rho} 
  \ee

%
%

 \subsection{The broken phase not too close to $T_{EW}$ }
In this case one can follow what has been done in the original
sphaleron  paper \cite{Klinkhamer:1984di}, substituting  into it appropriate 
couplings of the effective theory at finite temperature calculated in \cite{Kajantie:1995dw}.

We used the so called Ansatz B, expressing the sphaleron mass $M(R)$ and its r.m.s.
$\rho(R)$ versus its parameter $R$.  
Since this material is rather standard, we put the related expressions in the Appendix.
The results will be given in the next subsection.

\subsubsection{The sphaleron size distribution} \label{sec_sizes}


 We start with the unbroken phase, $T>T_{EW}$.
 The sphaleron size distribution in  can now be constructed using
 the mean mass (\ref{delta_M_v}), the small and large size limits
  (\ref{small_rho}) and  (\ref{large_rho}). More specifically, the distribution 
  interpolates between the small and large size distributions which merge
  at  $\rho=\rho_{\rm mid}=0.8\, {\rm GeV}$ to give (\ref{DMV}) 
  
 \begin{widetext}
\begin{eqnarray}
\label{PSIZE}
 P(\rho,T)\sim 
{\rm exp}\bigg[-{3\pi^2 \over g^2 T}\bigg({1\over  \rho} -{1\over  \rho_{\rm mid}}\bigg)\bigg]
\times {\rm exp}\bigg[- (0.457)^2\pi^2 g^2 T( \rho-\rho_{\rm mid})\bigg] \nonumber\\
\end{eqnarray}

%

\begin{figure}[h]
\begin{center}
\includegraphics[width=14.cm]{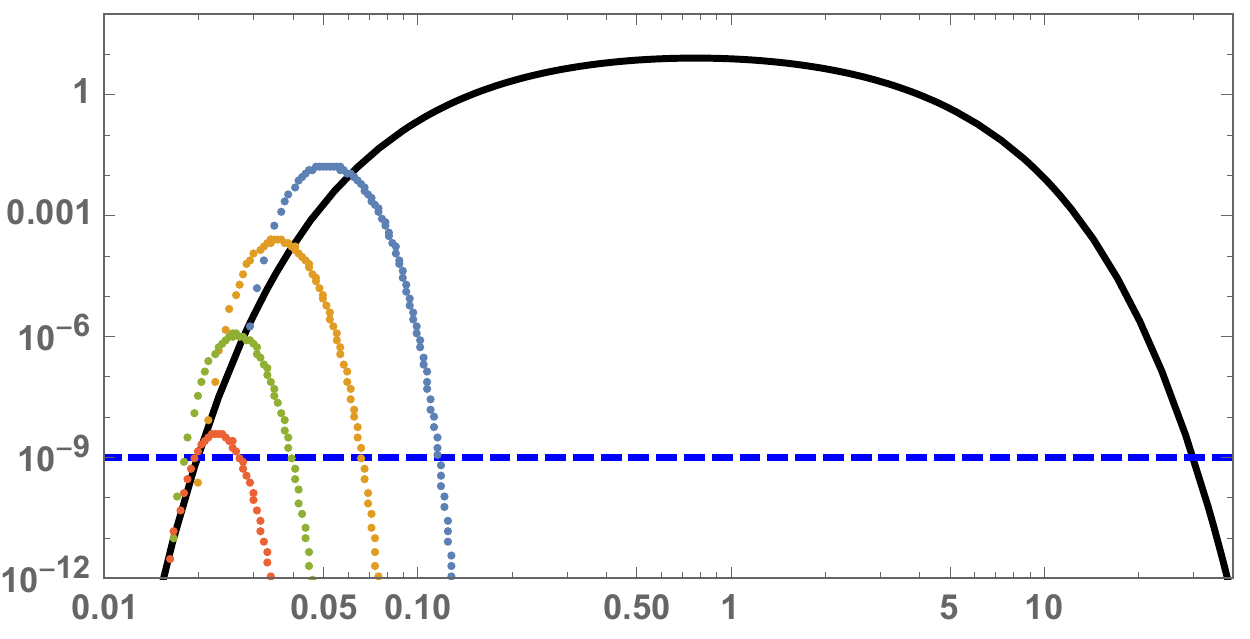}
\caption{The sphaleron suppression rates as a function of  the sphaleron size $\rho$ in ${\rm GeV}^{-1}$. 
The solid curve corresponds to the unbroken phase $v=0$ at $T=T_{EW}$. 
Four sets of points, top to bottom, are for well broken phase, at
$T=155,150,140,130\, {\rm GeV}$. They are calculated via Ansatz B described in Appendix C, and normalized to lattice-based rates. 
 The horizontal  dashed line indicates the Hubble expansion rate relative to these rates.}
\label{fig_P_of_rho}
\end{center}
\end{figure}
\end{widetext}

In Fig.~\ref{fig_P_of_rho} we show the sphaleron size distribution  at the
critical temperature (the solid line) and four temperatures below it, in the broken phase.
We see that the  appearance of a nonzero Higgs VEV leads not only
to a suppression of the rate, but also 
to a dramatic decrease 
of the sphaleron sizes. 
The lowest temperature shown,
$T_L\approx 130\,  {\rm GeV}$, corresponds to the sphaleron rate that reaches
 the Universe expansion rate (Hubble). 
 
 The intercept of each curve with  the dashed horizontal  line
gives (the smallest and) the largest size sphalerons which have rates comparable to the
Universe expansion rate, and should therefore be ``at freezeout", out of thermal equilibrium. 
One can see that for four sets of points shown, $\rho_{\rm max}$ changes from about $1/10\, {\rm GeV}$ to $1/30\, {\rm GeV}$. Very close to the critical temperature $T_{EW}$ the sphalerons may be
significantly larger in size, as seen from a comparison to the black curve.  However
the related uncertainty does not matter, as we will show below, because in this region the
CP asymmetry is extremely small, growing toward $T=130\, {\rm GeV}$.

\section{Sphaleron explosions: production of sound and gravity waves} 
\label{gravity_waves}

Most of the studies 
 on the gravity wave generation by the EWPT focus on scenarios based on the first order transition
 or  the ``cold" transition , as those usually yield large 
density fluctuations. To our knowledge, the smooth cross over transition of  the minimal SM 
 has not been considered. 

\begin{figure}[htbp]
\begin{center}
\includegraphics[width=6cm]{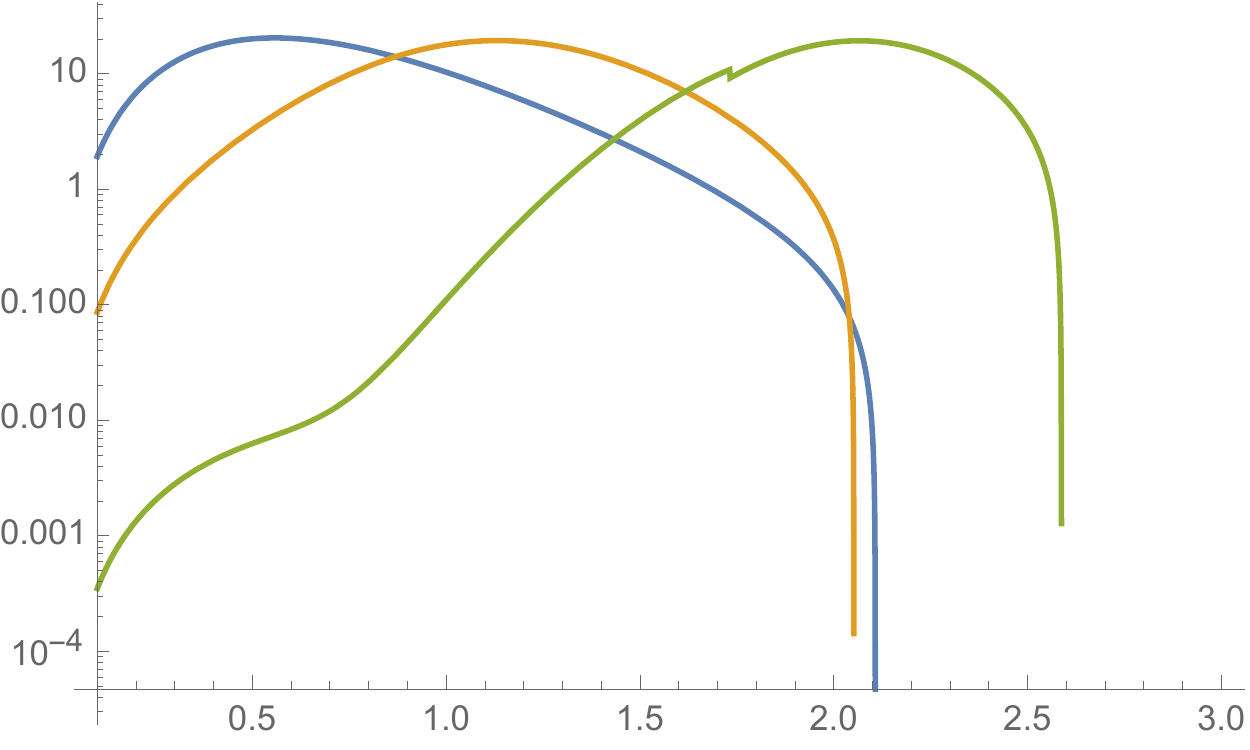}
\includegraphics[width=6cm]{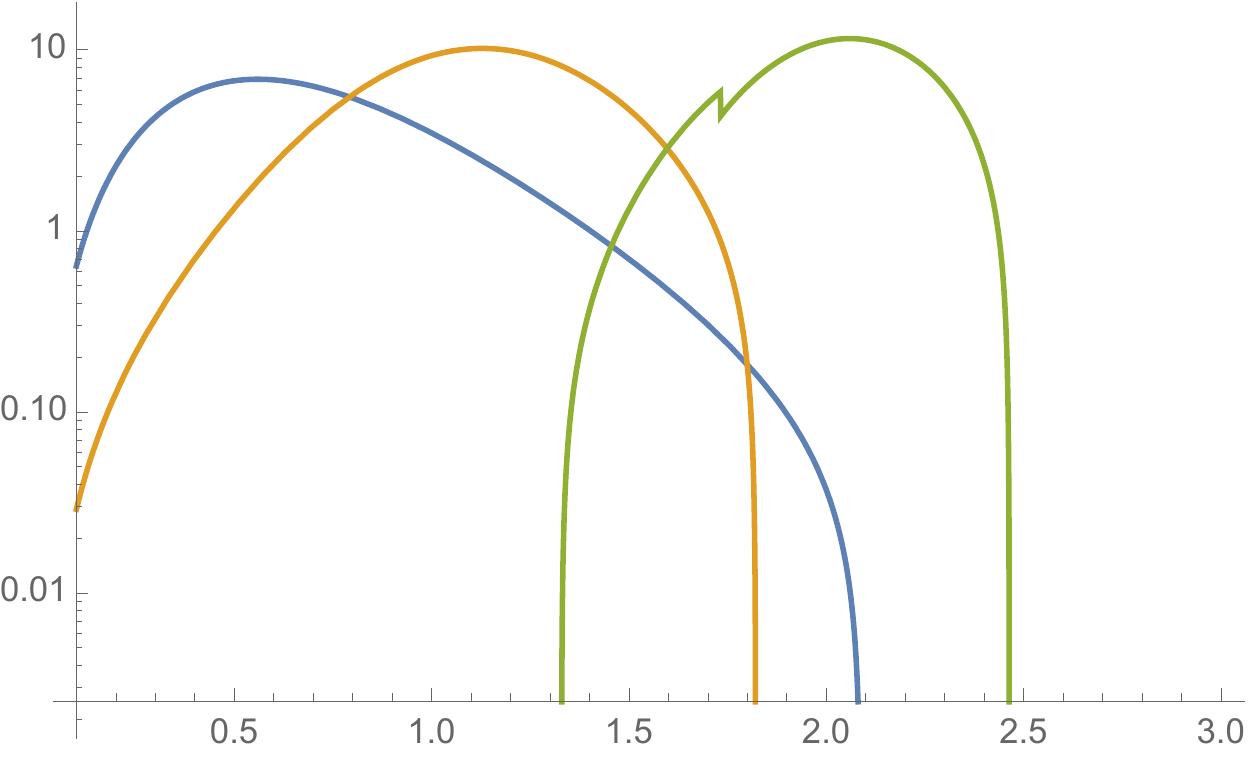}
\caption{Components of the stress tensor (times $r^2$, namely $r^2T^{00}(t,r)$ upper plot, $r^2T^{33}(t,r)$ lower plot) as a function of $r$, 
the distance from the center, at times $t/\rho=0.1, 1,2$, left to right.}
\label{figs_explosion}
\end{center}
\end{figure}

Since the sphaleron explosions give rise to significant deviations from a
homogeneous stress tensor of the plasma \be 
\Delta T^{\mu\nu}\sim G^{\mu\lambda}G_\lambda^\nu\sim { 1 \over g^4 T^4}
\ee
one may expect radiation of the gravity waves. The  stress tensor from the analytically known sphaleron field (\ref{eqn_field}) 
yields long expressions which are not suitable for reproduction here. Instead, we show 
in Fig.~\ref{figs_explosion} the behavior of $T^{00}(t,r)$ (the energy density) and  $T^{33}(t,r)$ (the pressure),
which illustrates the time-development of the exploding sphaleron in a spherical shell.

The key point here is to assess the scale dependence of both the sound and gravity waves 
triggered by the explosion, which can be expressed using the power-per-volume  $dE/d^4x$. 
Dimensional reasoning shows that  the average scale is shifted to smaller sphaleron sizes. 
The measure for small size sphalerons 

\be
{d\rho \over \rho^5} P(\rho)= d\rho\, {\rm exp}\bigg(-{3\pi^2 \over g^2 T\rho}-5{\rm log}(\rho) \bigg)
\ee
is peaked at 

\be 
\rho_*={3 \pi \over 20 \alpha_{EW} T} \approx {1 \over 10 \, {\rm GeV}  }
\label{rho_star} 
\ee
which is about an order of magnitude smaller than the peak of the distribution (Fig.\ref{fig_P_of_rho}). 

Also, for $T>T_c$ we do not expect direct gravitation emission from the sphaleron
explosion. In this regime the Higgs VEV vanishes, nothingh breaks the rotational
symmetry of the gauge field leading to spherically symmetric sphaleron explosions.
As a result, these explosions cannot {\it directly} generate gravitational waves no matter how violent
they are. This is not the case for $T<T_c$ as we discuss below. 

There is an {\em  indirect}  way to gravitational signal as discussed in~\cite{Kalaydzhyan:2014wca}. Spherical 
sphaleron explosions  do excite the underlying
medium through  hydrodynamical sound waves and vortices. Of course, 
the medium viscosity will eventually kill them, but since the damping rate scales 
 as $\Gamma\sim \eta k^2$,  at small $k$ (large wavelength) this time can be long. 
 Random set of sound sources creates 
 {\em  acoustic turbulence}. Under certain conditions it may 
turn into the regime of {\em  inverse cascade} and propagate many orders of magnitude,
perhaps to the infrared cutoff, the horizon size of the Universe.
It is a  $2\rightarrow 1$ generic process  \cite{Kalaydzhyan:2014wca} $$ {\rm sound+sound \rightarrow gravity \, wave} $$
which operates during the whole lifetime of the sound.

Just after the transition, at
 $T<T_{EW}$, a nonzero Higgs VEV leads to different masses
 of various quarks, leptons and gauge bosons. This ``mass separator" split
expanding spherical shell of the explosion into separate regions.
Also, a nonzero Weinberg angle (or the nonzero $g'$ coupling of the Higgs to the Abelian $U(1)$ field) produces an
 elliptic deformation of the sphaleron explosion.   It is created by the following part of the action \be 
\Delta S_a={m_Z^2-m_W^2 \over 2} \int d^4 x \sqrt{g} g_{\mu\nu}Z^\mu Z^\nu  
\ee
where the metric is explicitly shown. Writing it as a flat metric plus perturbation 
$g_{\mu\nu}=\eta_{\mu\nu}+h_{\mu\nu}$ and expanding in $h_{\mu\nu}$ is the standard way to derive the 
corresponding stress tensor, which is

 \be  
 \Delta T^{\mu\nu}_a= {m_Z^2-m_W^2 \over 2}  \bigg(-Z^\mu Z^\nu+{\eta^{\mu\nu}\over 2} Z^2 \bigg) 
 \ee
Here, the pre-factor is proportional to $v^2(T)$,  nonzero only after EWPT, at $T<T_{EW}$.

The power produced by the gravity wave is proportional to the 
 squared matrix element $|M|^2$ of the Fourier resolved stress tensor
by the gravity wave with momentum $\vec k$

\be 
{ \cal M}(h,k)= \int d^4 x \Delta T^{\mu\nu}(x) \,h_{\mu\nu} {  e^{ik\cdot x}\over r}   
\label{h_matrix_element_def}
\ee 
We recall that the polarization tensor for the gravity wave $h_{\mu\nu} $ is  traceless,  and transverse, i.e. 
nonzero only in the 2-d plane normal to $\vec k$.
For example, for $\vec k$ in the 1-direction, the pertinent contributions in (\ref{h_matrix_element_def})
are $T^{22}-T^{33}$ or $T^{23}$ for the respective polarizations. 

\begin{figure}[htbp]
\begin{center}
\includegraphics[width=7cm]{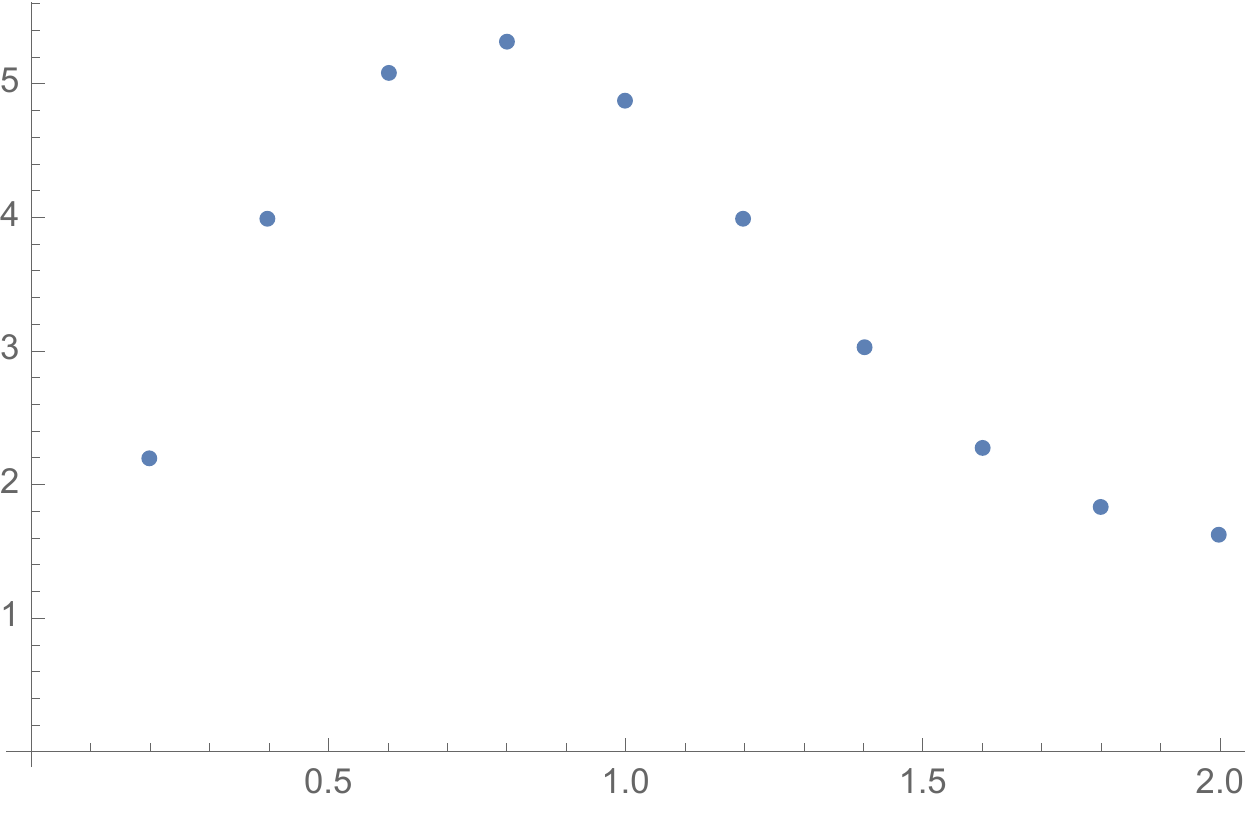}
\caption{The dimensionless matrix element ${\cal M}/(M_Z^2-M_W^2)^2$   in (\ref{h_matrix_element_def}) versus $k\rho$, 
for a gravity wave propagating in the 1-direction with transverse polarization giving $T^{22}-T^{33}$. }
\label{fig_h_matrix_element}
\end{center}
\end{figure}

The main part of the stress tensor gives vanishing matrix element, as it should, but the asymmetric 
part of the stress tensor produces gravitational radiation. In Fig.\ref{fig_h_matrix_element}  we show the dependence of the gravity wave matrix element as a function of $k\rho$. As expected, it is maximal at $k\rho \sim 1$. We have already evaluated the most important
sphaleron size in (\ref{rho_star}). As a result,  the expected 
gravitational wave momentum should be $k\sim 1/\rho_*\approx 10-30 \, {\rm GeV}$.

\section{CP violation and the sphaleron explosions} \label{CP}

\subsection{Standard Model CP violation, from the phase of the CKM matrix}

In this section we discuss whether the ``minimal" CP violation in the SM, 
following from the experimentally well studied complex contribution of the  CKM matrix, 
can generate  the required level of asymmetry.  Needless to say that this question 
was addressed by many in talks and textbooks, but it is worth reviewing it again here.


While the sphaleron decay process has been discussed at the classical level, 
see Appendix B, the CP-violating effects appear at one-loop level with the
contribution from   all generations of quarks and their interferences. 
We are not aware of any consistent calculation
of the corresponding CP violation during the sphaleron decays. 

Because of our focus
on large-size sphalerons, the appropriate strategy appears to be an 
evaluation of the fermionic determinant  in  the ``smooth" background of
a  W-field  with small momenta.  The determinant of the Dirac operator in such field 
${\rm Log(det}(\hat D))$ generates the effective action,
induced by one-loop fermion process, which is
similar to the well known Heisenberg-Euler effective action  in QED,  
with the CP-violating part extracted from its imaginary part.

Studies along these lines have been carried, but the  results are 
still  (to our knowledge) inconclusive. The calculation 
 in~\cite{Hernandez:2008db} found  no CP violation  to leading order ${\cal O}(W^4)$,
 but reported a nonzero contribution to order ${\cal O}(ZW^3 DW)$ from a dimension-6 P-odd and C-even operator 
 of the type

\begin{eqnarray}
\epsilon^{\mu\nu\lambda\sigma}
\left( Z_\mu W_{\nu\lambda}^+W_\alpha^-\left(W_\sigma^+W_\alpha^-+W_\alpha^+W_\sigma^-\right)+{\rm c.c.}\right)\nonumber\\
 \label{eqn_LCP}  
 \end{eqnarray}
 containing one neutral current vertex and the $Z$-boson field. 
 Other operators, which are C-odd and P-even, were claimed to contribute in \cite{GarciaRecio:2009zp,Brauner:2012gu}.
 
 It is not a trivial task to find an example of the field which would give a non-vanishing
 expectation value for this operator. In particular, it should be T-odd, and thus
 involving time evolution or electric field strength.
 We have checked using the analytic solution for the sphaleron explosion  
 \cite{Ostrovsky:2002cg,Shuryak:2002qz} described in Appendix B,  that it does give  a non-vanishing
expectation value  for this operator.  (The formulae are unfortunately too long to be given here.)

\subsection{Scale dependence of the CP violation} \label{CP_scale_dependence}

The main physical issue is not so much the operator expectation values, but rather the scale
dependence of the Wilsonian OPE coefficients multiplying them. These coefficients are
usually rather complicated  functions of the quark masses, see e.g. the
explicit form of the coefficient of the operator  (\ref{eqn_LCP})   given in the Appendix of \cite{Hernandez:2008db}. 
Instead of calculating the OPE coefficients for specific operators,
we suggest   a somewhat more general and universal approach, 
that will help us understand their scale dependence. 

Consider a typical CP 
violating contribution to the effective action in some smooth gauge background $W_\mu(x), Z_\mu(x)$
as vertices in a one-loop fermionic contribution. Each fermion line is 
characterized by a Dirac operator  $\Dslash$ in the gauge background. 
Using  left-right spinor notations it has the form

\begin{widetext}
\begin{eqnarray} \label{eqn_LR_operator} 
{\rm det}\left(  \begin{array}{cc} 
i\Dslash & M \\
M^\dagger  & i \partial\!\!\!/ \\
\end{array}
\right) 
= {\rm det}( i \partial\!\!\!/) \,{\rm det}\left(i\Dslash+M {1 \over  i \partial\!\!\!/ }M^\dagger \right)
\end{eqnarray}
\end{widetext}
where $M$ is a mass matrix in flavor space and  the slash here and below means the convolution with the Dirac matrices,
e.g. $\Dslash=D_\mu \gamma_\mu$. 
Let us use a representation in which this operator is diagonalized

\be 
i\Dslash  \psi_\lambda(x)=\lambda  \psi_\lambda(x) 
\ee
Its two sub-operators, $p\!\!\!/$ and $ W\!\!\!\!\!/$ are not in general 
diagonal in this basis, but for our qualitative argument we will only include their
diagonal parts

\be  
\left<\lambda |i \partial\!\!\!/| \lambda' \right> \approx p\!\!\!/ \, \delta_{\lambda \lambda'}, \,\,\,\,
\left<\lambda | W\!\!\!\!\!/ | \lambda' \right> \approx \xi \lambda\, \delta_{\lambda \lambda'} \ee
where $\pslash,\xi$ are in general some functions of $\lambda$. In this approximation 
the corresponding  (Eucidean) propagator  describing a quark of flavor $f$ propagating  in the background can be represented as the usual
 sum over modes 

\be 
S(x,y) \approx  \sum_\lambda {\psi^*_\lambda(y) \psi_\lambda(x) \over \lambda +M {p\!\!\!/}^{-1} M^+} 
\ee
where the right-handed operator $i\partslash$ 
is approximated by its diagonal matrix element in the $\lambda$-basis. Throughout, we will trade
the geometric mean appearing in all expressions
$\sqrt{p\!\!\!/\lambda} \rightarrow \lambda$, for simplicity.

The generic fourth-order diagram in the the weak interactions, shown in Fig.\ref{fig_5_diags}(0) contains   four CKM matrices.  
 In the coordinate representation its analytic form is form
\begin{widetext}
\begin{eqnarray}
\label{box}
 \int \prod_{i}^4 d^4 x_i  \, {\rm Tr}\bigg(W\!\!\!\!\!/(x_1) \hat V  \hat S_u(x_1,x_2) 
 W\!\!\!\!\!/(x_2)  \hat V^\dagger  \hat S_d (x_2,x_3) W\!\!\!\!\!/(x_3)  \hat V  \hat S_u  (x_3,x_4)
 W\!\!\!\!\!/ (x_4)   \hat V^\dagger   \hat S_d(x_4,x_1)\bigg)
\nonumber
\end{eqnarray}

\begin{figure}[h]
\begin{center}
\includegraphics[width=12.cm]{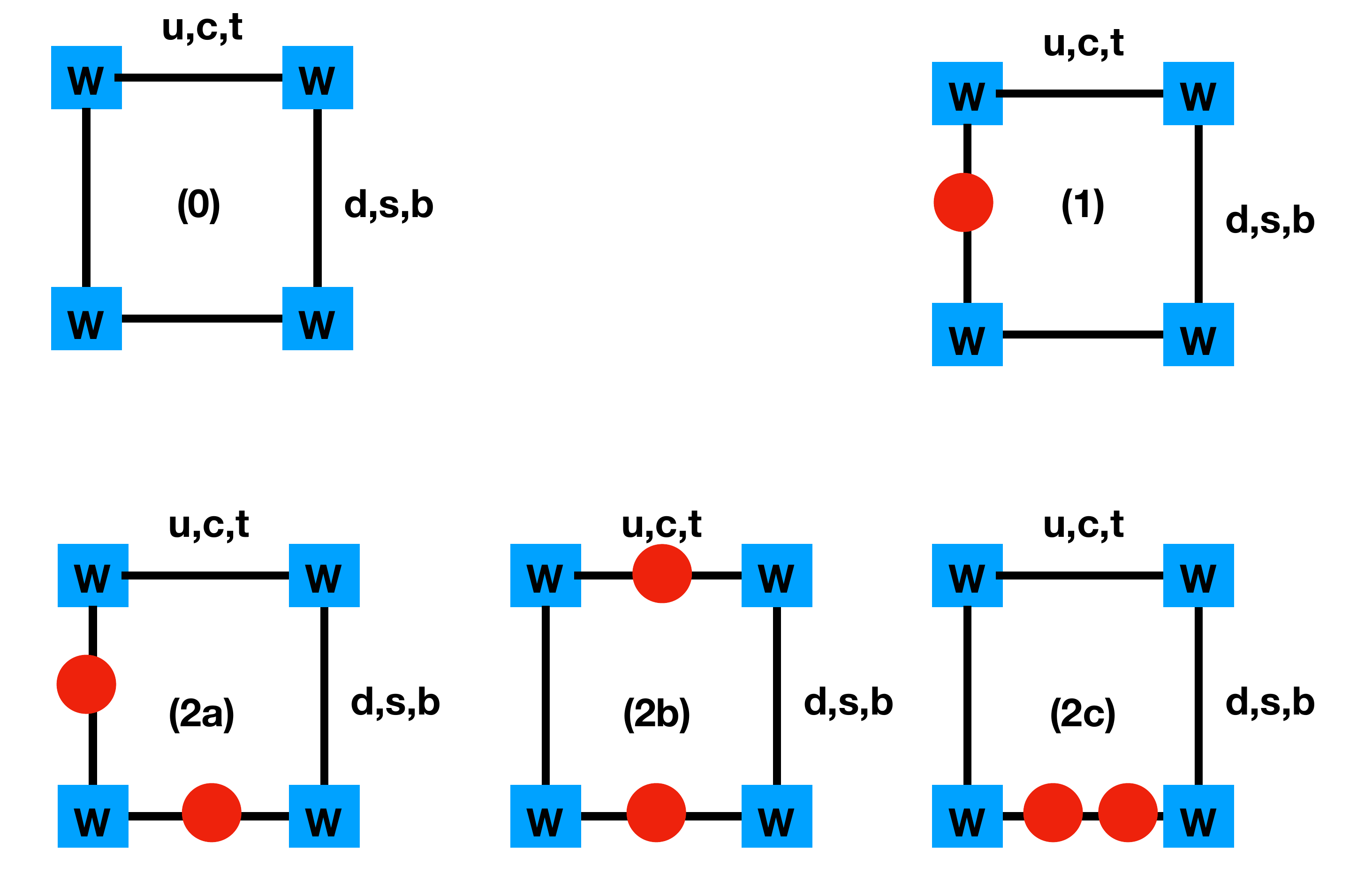}
\caption{Schematic description of a fourth order one-quark-loop diagrams with 4-$W$ boson insertions shown in blue squares.
The additional  $Z_\mu, A_\mu$ fields (diagram $(0)$), one $(1)$ and two ($2a,2b,2c$) are shown in red-circles. }
\label{fig_5_diags}
\end{center}
\end{figure}

\end{widetext}
Here hats indicate that the CKM matrix $V$  and the propagators
  are $3\times 3$ matrices in flavor space. The propagator also have
labels $u,d$, indicating up or down type quarks.   
The trace is over Dirac  and color indices as well: slash 
with $W$ field as usual mean convolution of vector potential with gamma matrices. 
For other diagrams of  Fig.\ref{fig_5_diags}  including $Z$ and electromagnetic $A^{em}$ fields  the analytic expressions are generalized straightforwardly.

The spin-Lorentz structure of the resulting effective action is very complicated. However, to understand the pattern of interference between quarks of different flavors, producing the $CP$ 
violation  and eventually its
scale dependence, one may focus on flavor indices alone. This is possible
if instead of standard $momentum$ representation of Feynman diagrams 
 we will use the {\em eigenbasis} of the Dirac operator in the background field. 
Specifically,  using the orthogonality condition of the different $\lambda$-modes 
and perform the integration over coordinates, to obtain  a simple expression, with a single sum over eigenvalues
$ \sum_\lambda F(\lambda)$ with 
 the same 4-th order box diagram  in the ``$\lambda$-representation" reduced to
\be
F(\lambda) =
\lambda^4 {\rm Tr}\left(  \hat V S_d  \hat V^\dagger  S_u  \hat V S_d  \hat V^\dagger  S_u \right)
\ee
 Note that different backgrounds have different spectra of the Dirac eigenvalues $\lambda$,
 and $F(\lambda)$ obtained should  therefore be convoluted with appropriate background-dependent
spectral densities $d(\lambda)$.

\begin{figure}[h!]
\begin{center}
\includegraphics[width=6.cm]{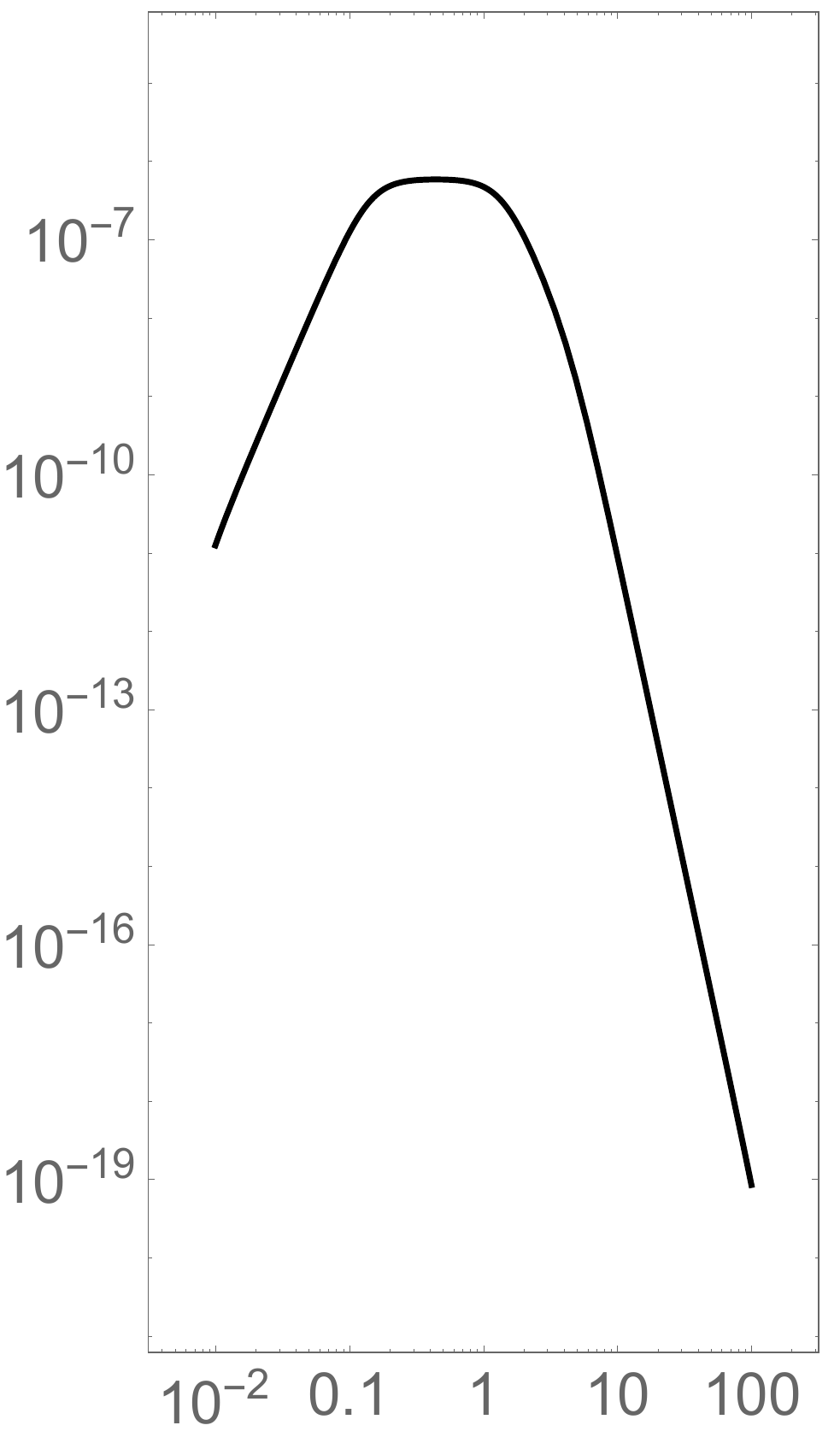}
\caption{CP-violating contribution $ {\rm Im} F_{ZZ}(\lambda)$  from the diagram $Z^2W^3DW$ versus $\lambda$ ({\rm GeV}).}
\label{fig_Fcp_lambda}
\end{center}
\end{figure}

However, in the eigenbasis representation 
 one can perform the multiplication of the flavor matrices
and extract  a $universal$ function of $\lambda$, that describes explicitly  the dependence of the  
CP violation contribution on a given scale.
Using the standard form of the CKM matrix $\hat V$, in terms of the
known three angles and the CP-violating phase $\delta$, and also the six known quark masses, 
one can perform the multiplication of these 8 flavor matrices and identify the lowest order  CP-violating term in the result.

After carrying  the matrix  multiplications in  the 4-th order diagram Fig.\ref{fig_5_diags}(0),  we arrive at 
a complicated expression, which when expanded  to the first power in the CP-violating phase 
does {\em not} have the $O(\delta)$ contribution. This means that 
the resulting effective Lagrangian  does $not$ produce CP violation.   
This conclusion is not new, as it was (to our knowledge) first obtained a decade ago in~\cite{Hernandez:2008db}. 

Since we do $not$ expect this cancellation between various quark flavors to hold for all 4-W insertions plus arbitrary 
neutral $Z$ and electromagnetic $A^{em}$ fields, we consider the lowest order insertions to the one-quark-loop as shown 
Fig.\ref{fig_5_diags} by red circles. The insertion of a Z-vertex is flavor independent and only double $\hat S\rightarrow \hat{ S}\hat{S}$ 
or triple $\hat S\rightarrow \hat{ S}\hat{S}\hat{S}$ the quark propagators as in diagram (2c). The electromagnetic 
field insertions carry quark electric charges,  $e_u=2/3$ and $e_d=-1/3$.


The explicit flavor traces for these higher order diagrams show that (1,2b,2c) do  not lead to CP-violating $O(\delta)$ terms in the resulting effective Lagrangian also.
The only one that does   is diagram (2a), with the result


\begin{widetext}
  \begin{eqnarray}
  \label{JAR}
&& {\rm Im} F(\lambda)=\lambda^6 {\rm  Im \,Tr}\left(  \hat V S_d  \hat V^\dagger S_u  \hat V S_d Z S_d  \hat V^\dagger  S_u Z S_u \right)\nonumber\\
 &&=2  \lambda^6 {J (m_b^2 - m_d^2) (m_b^2 - m_s^2) (m_d^2 - m_s^2) (m_c^2 - m_t^2) (m_c^2 - m_u^2) (m_t^2 - m_u^2) \over \Pi_{f=1..6} (\lambda^2 +m_f^2)^2 } 
 \end{eqnarray}
\end{widetext}
We recall that  $J$ is the famed Jarlskog
combination of the CKM cos and sin of all angles times the sin of the CP violating phase.

Note that the $x$-dependent gauge fields $W_\mu,Z_\mu,A^{em}_\mu$ in the vertices 
should be convoluted with the  currents in the eigenstates $\bar\psi_\lambda(x) \gamma_\mu \psi_\lambda(x)$. Diagram (2a) with 2-$Z$ and 2-$A^{em}$
would then carry  the extra factor

\be 
C_{2a}=\big((Z_\lambda)^2-{2\over 9} (A^{em}_\lambda))^2\big)
\ee
Since these two terms have opposite signs, no universal statement about
the $sign$ of the CP violation in arbitrary background can be made.

%

In Fig.\ref{fig_Fcp_lambda} we show (\ref{JAR}) as a function of the eigenvalue scale $\lambda$.
It is clear that  the magnitude
of the CP violation depends on the absolute scale very strongly. When the momentum scale
is at the electroweak value $\sim  100\,{\rm GeV}$ (the r.h.s. of the plot), it is $10^{-19}$. 
But  in the ``sweet spot", between the masses of the light and heavy quark  $\lambda \in (0.2-2)\, {\rm GeV}$,
the asymmetry is  suppressed only by about $\sim 10^{-6}$.

\section{Higher order plasma effects}

In the preceding section we discussed a generic effective Lagrangian in arbitrary background fields.
In fact we only need to focus on one particular background, that  originating from a {\em sphaleron explosion},
so we start by identifying its specific properties.

It is a $classical$ field, with $W_\mu,Z_\mu,A^{em}_\mu\sim O(1/g_{EW})$,
 and therefore the only smallness in the effective Lagrangian extracted from the one-quark-loop
 is $\sim O(g_{EW}^0)$, while the sphaleron action is  $\sim O(1/g_{EW}^{2})$. The perturbative treatment of 
 $W$ is related to the fact that we are looking for  a very small CP violation resulting from the CKM matrix. 
 The   perturbative treatment of $Z_\mu,A^{em}_\mu$ is  not a priori justified. It only helps in locating the lowest
 order nonzero diagrams.

The sphalerons and their explosion have so far been considered 
classically, via appropriate solutions of the Yang-Mills (and Dirac) equations.
The most symmetric case \cite{Shuryak:2002qz} (see Appendix) corresponds to symmetric phase $T>T_{EW}$ in which there is no Higgs VEV and the gauge fields are only the $SU(2)$
ones, without electromagnetic ones.   

Yet the actual sphaleron explosions happen in a primordial plasma. 
The generic argument is that the gauge fields are classical   $W,Z\sim O(1/g_{EW})$
while thermal fields of the plasma are   $\sim O(g_{EW}^0)$ , so in the leading order they are
not modified. However, since we consider sphalerons of different sizes, there appears
a parameter $\rho T$. On top of it, thermal fields have large number of degrees of freedom,
so the corrections to a classical approximation needs to be studied. 

The issue gets more urgent at the level of the one-loop quark-induced action we need to
study for CP violation. Quark fields, unlike gauge ones, are not classical. The eigenmodes
we consider are all normalized to a unit value. Plasma-induced polarization (mass) operators
come not only from weak interactions, but from thermal gluons as well.
Therefore the formal suppression parameter is $\sim \alpha_s$, which not so small.   
A full inclusion of all Dirac modes following a sphaleron explosion {\em in a plasma} is still beyond the scope  of this work.
In this section we provide a qualitative discussion and estimates.

\subsection{Electric screening}

Let us recall the argument put forth  at the beginning of section \ref{sec_sizes}:
since the sphalerons are {\em magnetic} objects,  they can be as large as allowed by the
{\em magnetic} screening mass, $$\rho \leq  {1 \over M_m} \sim {1 \over g^2 T} $$ Thus, 
it can differ from the basic thermal scale $T$ by about two orders of magnitude.
However, a sphaleron explosion generates an (electroweak) {\em electric} field as well.
Originally directed radially, that accelerates quarks and leptons from their initial 
state as zero modes, to their  positive and physical energy final states, 
violating baryon and lepton number.  In the final stage, only the 
transverse electric fields remain, producing physical $W,Z$ that are transversely polarized to 
the radial direction.  We recall  that under these conditions all particles are massless or have
small masses, as the Higgs VEV is zero or small.

The electric screening mass is $M_E\sim gT$, 
so the screening length is about $g$ times shorter than the magnetic screening length. The ``plasma on-shell masses" 
of $W,Z$ are also $M_{W,Z}\sim gT$. Both effects are incorporated by an additional 
thermal contribution to the effective Lagrangian (in momentum representation) 

\be 
\Delta L_{plasma}={1\over 2} \Pi_{\mu\nu}(T,k_\sigma) A^\mu(k_\sigma) A^\nu(k_\sigma) 
\ee
 with the one-loop polarization operator $\Pi$ calculated already in Ref.\cite{Shuryak:1977ut}.
 So, what qualitative modification these electric thermal effects produce?

Classically, the sphaleron explosion produces  $W,Z$ with momenta $p\sim 1/\rho$,
and their total number is of the order of the action $N_{W<Z}\sim 1/g^2\sim 100$. 
In the thermal plasma this is no longer possible. The available energy is not sufficient  to produce that many
gauge quanta since their thermal masses are  $M_{W,Z}\sim gT$. Furthermore, at the momenta 
$p\sim 1/\rho$ corresponding to the initial sphaleron sizes, the plasma modes are not 
$W,Z$ plasmons but rather collective modes of hydrodynamical origin, corresponding to
the longitudinal sounds (phonons) and transverse (rotational  and purely diffusive) motions.  
(How exactly the energy is divided between those modes we have not assessed yet:
it may be needed for gravity wave predictions). 
 
 As we already noted  earlier, due to the nonzero Weinberg angle, some part of sphaleron
 energy goes to QED electromagnetic fields and eventually to polarized magnetic clouds. Their polarization tensor includes not the
 electroweak but the electric coupling constant, which is smaller, and so their interaction with the plasma can probably be neglected, once they are produced.

 \subsection{How do plasma effects modify quark/lepton production and respective $B,L$ number violations  ? }

  In the primordial plasma, fermions are also modified by their interactions with  the thermal medium.
 While leptons have  electroweak interactions only,
  quarks  interact strongly with ambient  gluons,  with much a stronger coupling constant $g_s$. 
Therefore quark modes acquire  larger masses, and one may wander 
 if those can prevent the  $B,L$ number violation phenomenon itself.


At this point, a historical comment may be made.
When Farrar and Shaposhnikov~\cite{Farrar:1993hn}
realized that
 the CP violation induced by the CKM matrix have the ``sweet spot" mentioned above,
they focused  on the dynamics of the quarks with  {\em  small momenta} $p\sim 1 \, {\rm GeV} \ll T$. Specifically, they argued that under certain conditions 
the strange quarks  are totally reflected 
from the boundary of the bubble (they assumed the transition to be first order), while the up/down quarks are not. 
This scenario has been later criticized, based on higher order corrections to the quark
dispersion curves, and the conclusions in~\cite{Farrar:1993hn}, 
have been refuted.   For pedagogical reasons, let us split the refuting arguments in three, reflecting on their
increase sophistication.

{\em The first argument}  says  that the Euclidean thermal formulation with anti-periodic fermionic
boundary conditions, implies that the minimal fermionic energies are set by the lowest Matsubara mode

\be 
\omega_M= \pi T \sim 300\, {\rm GeV }
\ee  
Indeed the typical fermionic momenta  are of this order, and  the CKM-induced  CP violation at this scale is
$\sim 10^{-19}$ as we detailed above.  

 {\em The second argument} is based on  the emergence of a ``thermal Klimov-Weldon" quark mass

 \be M_{KW}={g_s T \over \sqrt{6}} \sim 50 \, {\rm GeV }  \label{eqn_MKW}  
 \ee
induced by the real part of the forward scattering amplitude of a gluon on a quark.   

Both  arguments were essentially rejected by Farrar and Shaposhnikov, who pointed to the fact  that while
both effects are indeed there,  there are still  quarks with small momenta $p\ll T, M_q$  in the Dirac spectrum.

{\em The third argument} which is stronger, 
was given in~\cite{Gavela:1993ts,Huet:1994jb}.  It is based on the
{\em decoherence} suffered by a quark while traveling  in a thermal plasma, as caused by the  imaginary part
of the forward scattering amplitude (related by unitarity to the
cross section of {\em non-forward} scatterings on gluons). Basically,  they argued that if a quark starts with a small momentum,
it will not be able to keep  it small for necessary long time, due to such scattering.  The imaginary part is about

\be {\rm Im} (M_q) \sim \alpha_s T \sim 20\, {\rm GeV }
\ee

We now return  to the sphaleron explosions we have presented, and ask how such plasma effects
can affect their quark production. 
The most obvious question is that of ``insufficient energy". 
Indeed, if each quark carries a  ``thermal Klimov-Weldon mass" as the smallest energy at small momenta,
is there even enough energy to produce the
expected 9 quarks? Alltogether, these 9 masses 
amount  to about $450$ GeV,  which is  comparable to the total sphaleron mass  (\ref{small_rho}) at 
a size $\rho\sim 1$ GeV$^{-1}$.  
Therefore, the classical treatment used above, in which the back-reaction of the quarks
on the explosion were neglected, by solving the Dirac equation in a background field approximation,
should be significantly modified.

However, there is a  simple  way around the ``insufficient energy" argument.
 In thermal field theory the sign of the imaginary part of the 
 effective quark mass operator can be both positive or  negative. This corresponds to
 the fact that instead of producing new quarks, the sphaleron amplitude can
 instead {\em absorb}  thermal {\em antiquarks} from the plasma.

Still we would argue that, unlike the Farrar-Shaposhnikov scenario~\cite{Farrar:1993hn}, 
our sphaleron-induced baryon number violation should survive all plasma effects. 
We do not classify quarks by their momenta, but rather by the virtualities or {\em eigenvalues} 
 of the Dirac operator $\lambda$, in the background of sphaleron explosion solution.
 
 It is true that the plasma effect will modify the spectral density $P(\lambda)$ in a
 way, that most of the virtualities $\lambda$  are equal or larger than the thermal Klimov-Weldon mass
 (\ref{eqn_MKW} ) $$|\lambda| > M_{KW}$$ According to Fig.~\ref{fig_Fcp_lambda}, this puts the
 CP asymmetry to be of order $\sim 10^{-17}$, way too small for BAU.

However, a sphaleron explosion is
a phenomenon in which  {\em  gauge topology} of the background field is changing. The
topological theorems requires the  existence of a 
 {\em   zero mode} of the Dirac operator in the spectral density, $P(\lambda)\sim \delta(\lambda)$. 
 (Another way to say it, is to recall  that the sphaleron explosion  implies
changing of the Chern-Simons number, which is locked to the change in 
the quarks and leptons  left-polarization by the  chiral anomaly.)  
Plasma effects do indeed modify the gauge fields during the sphaleron explosion,
 perhaps strongly, $O(1)$ in magnitude, but 
 they cannot  change their {\em topology}. Therefore plasma effects cannot 
 negate the existence of the zero mode: it is robust, completely {\em immune to perturbations}.

%
%

For a skeptical reader, let us provide an example from practical lattice gauge theory simulations,
which may perhaps be convincing. At temperature at and above
the critical $T>T_c$, in a QGP phase, there are plenty of thermal gluons. 
And yet, when a configuration with the topological charge $Q=\pm 1$
is identified on the lattice, an exact Dirac eigenvalue with $\lambda=0$ is observed,  within 
 the numerical accuracy, typically $10^{-9}$ or better. Also, the spatial shape of this eigenvalue is in very good agreement
 with that calculated using semiclassical instanton-dyons~\cite{Larsen:2018crg}.  Zillions of
 thermal gluons apparently have no visible effect on the shape of these modes, in spite of the fact
 that the gauge fields themselves are undoubtedly strongly modified.  
 Of course, this example is in an Euclidean time setting, while the sphaleron explosion is
 in  a Minkowskian time setting. Real time simulations are much more costly and have 
 only been done with gauge fields {\em without} fermions.  However, we are confident
 that baryon number violation itself is completely robust, immune to thermal modifications.

 \subsection{Dirac zero mode and related CP violation}

Now that we argued  that the Dirac operator should still have an exact zero mode for a sphaleron explosion,  even in the plasma, we now
 further ask how its presence in the Dirac operator determinant can affect
 the estimates of the CP violation we made earlier.

 \subsubsection{Dirac zero mode}

 Let us return to the Dirac operator (\ref{eqn_LR_operator} ), in left-right notations.
The quark-gluon scattering is vectorial, so
 the Klimov-Weldon mass (or more generally, the forward scattering amplitude at an appropriate momentum)  should  be added 
 to the (LL) and (RR) diagonal elements
 
 \be
 {\rm det}\left(i\Dslash+M_{KW} +M_{LR} {1 \over  i \partial\!\!\!/ +M_{KW}}M^\dagger_{RL} \right)\nonumber\\
\ee
Note that the non-diagonal fermion masses (LR and RL) flipping chirality
can only come from interaction with the Higgs scalar field, violating chiral symmetry. 
When there is no Higgs VEV (at $T>T_c$), to lowest order the last term is absent. At next
order,  it is proportional to the corresponding Yukawa couplings for different fermion
species. 

What we argued above, means that the plasma-deformed 
 first LL operator $i\Dslash+M_{KW}$ should, like the vacuum version,
 have a zero eigenvalue. For that, we  write

\be
 i\Dslash= ( i \partial\!\!\!/ +g A_\mu )\hat 1 +g A_\mu (\hat M_{CKM} - \hat 1) \label{eqn_Dslash2}
 \ee
where hats indicate matrices in quark flavors. The topological zero
mode $\lambda=0$ follows from the flavor-diagonal part, as a zero eigenvalue of
the first bracket
\be  
(i \partial\!\!\!/ +g A_\mu \hat 1 )\psi_\lambda=\lambda \psi_\lambda 
\ee
The so called ``topological stability"  implies that the zero eigenvalue  {\em does not}
have any perturbative corrections.

 The remaining part of the Dirac operator (\ref{eqn_Dslash2}) can formally be considered small
 and thus treated perturbatively, providing small modification of the known explicit solution to the Dirac equation in the background of  sphaleron explosion.
  The deviation of the gauge field term from $\hat 1$ through
 $$
g A_\mu (\hat V_{CKM} - \hat 1)$$
 would provide vertices for flavor changing quarks, and the
mass term, in the form $$ M_{LR} {1 \over  i \partial\!\!\!/ +M_{KW}}M^\dagger_{RL} $$
would provide perturbative corrections to the quark propagators connecting these vertices.
As we will see, this  flavor-dependent part is key for evaluating the magnitude of CP violation.

\subsubsection{Quark production probability}

As explained by 't Hooft long ago~\cite{tHooft:1976snw}, the physical meaning of the zero mode of the instanton 
(or its analytic continuation to Minkowski time~\cite{Shuryak:2002qz} we imply here) is the wave function 
of the outgoing fermion produced. 
The CP violation induced by the quarks ``on their way out" appear due to interferences of certain diagrams 
with different intermediate states. In short, the production
probabilities of quarks and   antiquarks are {\em not} equal. The method to calculate
the effect was previously developed by Burnier and one of us~\cite{Burnier:2011wx}.

Consider an outgoing  quark, accelerated by the electric field of the sphaleron explosion, 
 and  interacting on its way   with the $W$ field {\em twice}. The full probability for the quark production 
contains sums over all possible intermediate flavor states. For example, if the quark started as
$b$-quark, then one has a triple sum over intermediate flavors

\begin{widetext}
\begin{eqnarray}
P_b=\sum_{IJK}  \bigg(A\bigg(b \rightarrow I=t,c,u \rightarrow J=b,s,d\bigg)\times
A\bigg(J=b,s,d \leftarrow K=t,c,u  \leftarrow b\bigg)\bigg)
\end{eqnarray}
We now note  the three key features of this expression:\\\\
(i) the intermediate up-quarks $t,c,u$ in each amplitude need not be the same.
The interference of multiple paths in flavor space, induced by
the CKM matrix angles, may lead to CP violation;\\
(ii) the total number of CKM matrices $\hat V_{CKM}$ is four, which is just enough to make
this CP violating contribution nonzero; \\
(ii)  the combination $\hat V_{CKM}^+ \hat V_{CKM}\hat V_{CKM}^+\hat  V_{CKM}$ and  its complex conjugate is 
not the same as for the corresponding ($\bar b$) antiquark. \\\\
In light of this, 
the probability to produce a quark and an antiquark are not equal, i.e. $AA_q\neq \overline{AA_{q}}$.
More specifically, let us write the convolution for a particular initial up-quark state labeled as $U0$,   

\begin{eqnarray}
\label{AAU1}
AA_{U0} \sim &&\sum_{D1,U,D2}{\rm Tr } \bigg(\hat P_{U0}  W(x_1)\hat V_{CKM} S^{D1,D1}( x_1,x_2) \nonumber\\
\times&& W(x_2) \hat V_{CKM}^+  \tilde  S^{U1,U1}(x_2,x_3) W(x_3)\hat V_{CKM}
S^{D2,D2}(x_3,x_4)W(x_4)\hat V_{CKM}^+  \hat P_{U0} \bigg)
\end{eqnarray}
where $\hat V_{CKM}$ is a $3\times 3$  CKM matrix (indices not shown)
and $P_{U0} $ at both ends  are projectors onto the original quark type.
The propagators $S(x,y)$ are diagonal flavor matrices with their
indices shown. There are 3 options for  each index, 
$D1,D2=b,s,d,$ and $U0,U1=t,c,u$, so for each quark the probability has $3^3=27$
interfering terms. The intermediate propagator 
$\tilde S^{U1,U1}$ has tilde, which indicates that it should include the propagation from
point $x_3$ to infinity, and its conjugate propagation from infinity to point $x_4$. Therefore,
its additional phase depends on the distance between these points.   
The corresponding amplitudes for the 
antiquarks involve complex conjugate (not Hermitian conjugate!) CKM matrices
relative to the quark amplitude, namely

\begin{eqnarray}
\label{AAU2}
 \overline{AA}_{U0} \sim &&\sum_{D1,U,D2} Tr \hat P_{U0}  W(x_1)\hat V_{CKM}^*  S^{D1,D1}( x_1,x_2) \nonumber\\
&&\times W(x_2) \hat V_{CKM}^T  \tilde  S^{U1,U1}(x_2,x_3) W(x_3)\hat V_{CKM}^*
S^{D2,D2}(x_3,x_4)W(x_4)\hat V_{CKM}^T  \hat P_{U0} 
\end{eqnarray}
\end{widetext}
The difference in the probability of production of a quark and antiquark is denoted by
$$ \Delta P_Q\equiv {\rm Im}\big(AA_{Q} -\overline{AA}_{Q}\big) $$ 
We now note that:\\\\
 (i) the propagators of quarks of different flavors between the same relative points 
have different phases; \\
(ii) the locations in the amplitude  $x_{1,2}$ need not be
the same as  the locations $x_{3,4}$ in the conjugate amplitude, so  in principle  we need
to integrate over all of these locations independently.\\

For a qualitative estimate of (\ref{AAU1}-\ref{AAU2}) we write the 
nontrivial flavor-dependent phases in the propagators as
$$ S^{QQ}=e^{i\phi_Q} $$ suppressing for an estimate their dependence on the coordinates,
and perform the sums with $U0$ referring to all 6 initial types of quarks. The lengthty result is
given in Appendix C. As already indicated, these phases 
come from the  last term in the Dirac operator. Apart of common phase induced by the $\pslash$ in it, there are flavor-dependent 
phases induced by the last term in the Dirac operator
\be 
\phi_Q = {m_Q^2 |x_1-x_2| \over M_{KW} } 
\ee
Using for coordinate distance travelled the sphaleron size (maximal at freezout line)
 $$|x_1-x_2| \approx \rho_{max} \sim 1/10-1/30 \, {\rm GeV}^{-1}$$
we introduce a new  (temperature-dependent) mass scale
\be 
M_\rho\equiv \left ({M_{KW} \over \rho_{max}(T) }\right)^{1/2} \sim 40\, {\rm GeV }
\ee 
Using this notation, the additional phases is just a ratio of 
(flavor and temperature-dependent) quark mass to  $ M_\rho$, squared: 
\be 
\phi_Q = {m_Q^2  \over M_\rho^2} 
\ee

  When the quark masses are smaller than $ M_\rho$,
the corresponding phases are small.

\subsubsection{Amount of CP violation}

Let us now recall that we are discussing
the  Universe at temperatures  across the electroweak transition, with the  Higgs VEV  $v(T)$ emerging from zero to eventually its value in the broken phase
as we have it today. For a specific expression see the  lattice result (\ref{lattice_v_of_T}).
 All  quark masses grow in proportions to the VEV, and therefore the ensuing CP violation grows.
 We will divide this stage of the evolution into  two stages.

{\em Stage 1:}
In the quark production probabilities, the four vertices with CKM matrices are connected by three propagators, leading
to expressions cubic in  $\phi_Q\sim (m_Q/M_\rho)^2$, after expanding 
the expressions in Appendix C. The end of stage 1 happens when the largest of the
phases, that due to the top quark, reaches $O(1)$, or
\be m_t \approx M_\rho \sim 40\, {\rm GeV} \ee 
At this time  all other quark masses are much smaller than the top quark mass,
respective to their Yukawa couplings, and their 
phases are therefore  small. 
The lengthy expressions in Appendix C  can be 
simplified by expanding these exponents to  first order in the phases. Say, the one for $d$
quark contains the heaviest quark masses in the expression, and the corresponding  CP  asymmetry is

\begin{widetext}
\be 
\sim 2J {m_b^2(T) m_c^2(T) \over M_\rho^4(T)} \sim 2J \bigg({m_b^2(0)  m_c^2(0) \over m_t(0)^4 }\bigg)\bigg({m_t(0)^4 \over M_\rho^4}\bigg)
\approx J \cdot 1.2\cdot 10^{-7} \cdot 350\sim 10^{-9} 
\ee
\end{widetext}

{\em Stage 2:}
This  corresponds to  a {\em large} top quark mass $m_t>M_{\rho}$ and the phase $\phi_t \gg 1$, with a 
 rapidly oscillating exponent. Therefore, we  assume 
$$ e^{\pm i \phi_t } \approx 0 $$
and drop all factors with top quark phase. If one starts from a light quark $U0=u$, the  resulting 
expression contains the mass differences with the heaviest remaining masses of $b,c$
quarks, namely

\be 
\label{RATIOX}
2 J {(m_b^2-m_s^2)(m_c^2-m_u^2) \over M_\rho^4} 
\ee
It is similar to the expression we had before, but with masses continuing to  grow.
The numerator grows as the fourth power of VEV $\sim v(T)^4$, and the denominator
approximately as its second power due to sphaleron size shrinkage. 
As a result, the temperature dependence is $\sim v^2(T) \sim (T_{EW}-T)$.

Eventually, the temperature falls to $T=130\, GeV$ below which the sphalerons freezeout
completely.  The  prefactor $2J\sim 6\cdot 10^{-5}$ and 
 the CP asymmetry 
(\ref{RATIOX}) is about  \be A_{CP}\sim 0.25\cdot  10^{-9} \ee 
comparable to what one gets by the end of stage one.

Some remarks are now in order here.  Note that if one starts with the first generation $u$ quark, the intermediate ones kept are $b$ and $c$, of the third and second generations. So, as required, all three generations are involved. Yet this does not mean that {\em all 6 quark flavors} 
need to be involved: in particular the answer  $\Delta P_u$ (in Appendix D) contains  a factor $(m_b^2-m_s^2)$ but not  $(m_d^2-m_s^2)$, as there is no $d$ quark anywhere.  Thus there is no $m_s^2$ in our answer.
The situation is exactly the same as in the exclusive $b$ decays as we discussed earlier. Only the masses of the quarks explicitly involved
in the process, not all 6 mass differences,  needs to be present. The full Jarlskog mass factor is not required in exclusive reactions. 

Yet the symmetry between quarks strikes back: the CP violation for  the $d$ quark,
 $\Delta P_d$  (in Appendix D), has the same magnitude of the CP violation (\ref{RATIOX}) but have {\em  the opposite sign}. Therefore, in the symmetric phase at $T>T_{EW}$,
 when orientation of the sphaleron zero mode in $SU(2)$ group space is spherically symmetric, one has {\em cancellation of the CP violations}, between contributions of sphalerons which produce more $u$ or more $d$ quarks. Such cancellations
 is similar to what is seen in leading order effective Lagrangians, and they are is expected to be violated if
 higher order effects, e.g. including electromagnetic interactions, are taken into account.
 
As the  $SU(2)$ symmetry gets broken at $T<T_{EW}$ (the phase we discuss), there is  no more any symmetry between up and down weak isospin orientations. Specific Lagrangian for quark interaction with $Z$ field takes the well known form
\be L_{\bar q q Z}=-{g \over 2 cos(\theta_W)} \sum_i \bar q_i Z_\mu \gamma^\mu (g^i_V-g^i_A \gamma^5) q_i \ee
in which vector and axial constant are different for up and down quarks:
$$g_V^i=t(i)-2Q_i sin^2( \theta_W) ,\,\,\, g_A^i=t(i) $$
where $t(\pm)=\pm 1/2$ is weak isospin and $Q_i=(2/3,-1/3)$ are quark electric charges.
 Therefore the $u$ and $d$ CP-violating terms do $not$ cancel each other. 
 Since $Q_u-Q_d=+1 $ and  sine of the Weinberg angle ${\rm sin}(\theta_W)\approx \frac 12 $
 are both of order 1,
 the effective CP violation in the sphaleron explosion 
remains of order $10^{-9}$.
 
(The exact magnitude and, most importantly, the absolute sign of 
baryon number produced, require calculation of the convolution of quark zero mode
with background $Z$ field of the sphaleron explosion, which is beyond the boundaries of this paper.) 

%
%
%

Completing this section, let us recapitulate the assumptions made, and provide additional comments on further steps of this program:
1/ we used the sphaleron size  $\rho$ as  a placeholder for the distance between points at which the fields appear. In real calculation, coordinates
should be integrated over with projections to currents, in an actual Feynman diagram
defined on top of the fermionic zero mode of sphaleron explosion;
2/  we eliminated  the term with the largest
phase, that with the top quark mass, 
assuming that the oscillating term leads to the cancellation of all terms and zero answer.
This contribution can be studied  further; 

Note that  we are discussing the cosmological time near the phase transition, the 
 quark masses under consideration are not fixed but vary in time, from zero to their physical
 values in the broken phase, due to a changing
Higgs VEV $v$. Suppose we  consider the situation in which
the largest  phase   is of order one. We estimate this to happen when 
 \be m_t(T)   \sim \sqrt{E/\rho} \sim M_\rho \label{cond_top} 
 \ee
 This delicate estimate is straightforward in logic, but relies on a key assumption,
 namely that  the energy
 of the outgoing quark is larger than $E\sim M_{KW}$. Naively, it  cannot be smaller
 for free quarks in the electroweak plasma, as their interaction with the gluons
 makes their energy of order $M_{KW}$ even at zero momentum. Yet this assumption can
 be amended by the fact that the outgoing quarks are {\em not} free.  They are still
 in the sphaleron field where it is worth recalling that they satisfy
 
\be
\big(-D^2+g^2 G_{\mu\nu} \sigma^{\mu\nu}\big)\psi_\lambda=\lambda^2   \psi_\lambda
\ee
 which contains not only the momentum squared, but also the $g^2 A^2, g^2 G_{\mu\nu}$ term.
For a sphaleron, the latter is about $\sim 10/\rho^2$, with the sign
  depending on the location. Therefore, the eigenvalue spectrum does not start
 at $\lambda>M_{KW}$ sharply, but extends to smaller values. Indeed,
  the quarks produced are pulled from the lower continuum, or the Dirac sea.The
  numerical value estimated above contains $1/\lambda^4$, and thus the result
  depends on the tail of the spectral density at smaller $\lambda$.

%

\section{Baryogenesis} \label{BAU}

\subsection{Which sphaleron transitions are out of equilibrium?} \label{freezeout}

Before we discuss freezeout of the sphaleron transitions, it is instructive to recall
an analogous case of freezeout of the  ``little Bang" in heavy ion collisions. 
A good example is the  production of antinucleons $\bar N$. 
In the 1990's the cascade codes predicted small yield of $\bar N$, based on the fact that 
on average many baryons surround an anti-nucleon. Since the annihilation cross section
$\sigma_{N\bar{N}}$ is large, the anti-nucleon  lifetime $\tau\sim 1/(n_N \sigma_{N\bar{N}} \langle v \rangle)$ must be quite short. 
However, the data showed otherwise, with a number of produced anti-nucleons much larger than predicted by the numerical codes.
The explanation was given in~\cite{Rapp:2000gy}. The annihilation
creates multi-pion final states  with $N_\pi\sim 6$, and the inverse reaction $N_\pi\rightarrow 
N\bar{N}$ was ignored because of certain prejudice, that the multi-particle collision
has negligible rate. Explicit calculations showed otherwise, in agreement with detailed balance in thermal equilibrium. 

This equilibrium is only violated after the so called {\em  chemical freezeout}, 
 when  the rate $\Gamma_{inelastic}$ of the inelastic reactions changing $N_\pi$ and $N_N$ 
gets {\em smaller} than the expansion rate of the fireball $H=\partial_\mu u^\mu$ (the Hubble of the
Little Bang). While the particle numbers become time independent, 
 the thermal state of the expanding fireball is described via time-dependent chemical potentials,  $\mu_\pi(t)$ and $\mu_N(t)$. 
The annihilation channel contains the fugacity factor ${\rm exp}(-2\mu_N/T)$, while the inverse
reaction channel contains the fugacity ${\rm exp}(-N_\pi \mu_\pi/T)$. Since

\be  N_\pi \mu_\pi > 2\mu_N 
\ee
the inverse production process gets more suppressed than the direct  annihilation process. 
Only then,  the anti-baryon population starts to be somewhat depleted.

We now return to Sakharov$^\prime$ s conditions for BAU, the deviation from thermal equilibrium. 
The sphaleron transitions  basically consist of two different stages. The first is a 
complicated diffusion of the gauge fields {\em moving uphill} (say from $N_{CS}=0$ to $\frac 12$)  by thermal fluctuations, driving the fields to
the sphaleron configuration at $N_{CS}=1/2$. Fortunately, we  do not need to understand
it. In equilibrium the sphaleron population is given by the Boltzmann factor ${\rm exp}({-E_{\rm sp}/T})$.
The second  is the sphaleron decay  {\em  rolling downhill} , say from $N_{CS}=\frac 12$ to 1,   as described by the real-time solution of the
equations of motion in Appendix C.  The process is purely Minkowskian with an amplitude $e^{iS}$ and a real action, hence unit probability 
of realization.

In equilibrium,  the principle of the detailed balance requires that the inverse reaction
with $t\rightarrow -t$, has the same rate. It means that, contrary to prejudice it may still
take place, where a  large number of gauge quanta $$N_W\sim 1/\alpha_{EW}\gg 1$$ 
%
plus the 12 fermions required by the anomaly relation, can collide together, putting the field
back on top of the sphaleron hill.  As Sakharov argued, the presence of CP and thus T-violation 
in the process matrix element does not matter. Thermal  occupation factors depend only on masses/energies, 
which are  CP invariant.


However, since the Universe is expanding at a  Hubble rate $H\sim 1/t_{EW}$, 
some of these transitions involve  particle changing  rates smaller than the Hubble
rate. They are  out of equilibrium! Earlier, we have shown that as a function of the 
sphaleron size $\rho$, the sphaleron decays get frozen when

\be 
{\rm exp}\bigg(-B_{\rm sph} \rho v^2(T)\bigg) < 10^{-9} 
\ee
We now argue that the inverse process is frozen differently,
so that the  the equilibrium condition and its detailed balance become violated.

More specifically,  for the large-$\rho$ tail and in the  small-$v$ regime near $T_c$, the {\em inverse} reaction of multi-quanta
 collisions gets frozen first. The argument is based on the observation that the corrections to the sphaleron mass 
 $\Delta E_{\rm sp}=C_{\rm sp} \rho v^2$ is smaller than the modification of the thermal Boltzmann factor of the inverse reaction.
 The latter can be written as corrections to ultra-relativistic energies of W bosons due to their mass
 $  E_p\approx  p+ M_W^2/2p$, so the energy in their thermal exponent  changes by
 
 \be 
 \Delta E_{W}=\sum_{i=1}^{N_W}\Delta E_i\approx \frac {N_W}{2p}\bigg(\frac {M_W(0)}{v(0)/v(T)} \bigg)^2
 \ee 
 after rescaling the W-mass.
Since $1/p\sim\rho$, this correction is of order $\sim \rho v^2$, but the coefficient $N_W$ is parametrically  larger 
with  $N_W\sim 100 ={\cal O}(1/\alpha_{EW})$.

\subsection{Contribution to BAU from out-of-equilibrium sphalerons } 
\label{BAU_estimate}

Our next step is to calculate the BAU  produced by the large-size sphalerons
which are out of equilibrium. As detailed above, this requires moving to the
freezeout point, thereby sacrificing 9 orders of magnitude in the rate with
$F_{\rm freezeout}\sim 10^{-9}$. This is the regime
where the sphalerons decay but are not regenerated.
Each electroweak sphaleron changes the baryon number by 3 units , i.e. 
 9 quarks  each carrying $\frac 13$ baryon charge. The baryon number density 
 normalized to the entropy density of  matter, follows by integrating the rate over the 
 freezout time  $\Delta t_{FO}$
 
 \begin{widetext}
    \begin{eqnarray}
    \label{ENT}
   \bigg({n_B \over s}\bigg) = 3 A_{CP}\times \bigg[{\Gamma\,F_{\rm freezeout} \over T _{EW}s_{EW}}\bigg]\times 
   \bigg[T_{EW} t_{EW}\bigg] \times \bigg[{ t_{FO}- t_{EW} \over t_{EW}} \bigg]
   \end{eqnarray}
   \end{widetext}
Here  $A _{CP}$ is the CP asymmetry, the relative difference between baryon number production and annihilation
 in a single sphaleron transition.   The second factor in square bracket is  the out-of-equilibrium sphaleron rate  normalized 
 to the total entropy density $s_{EW}/T_{EW}^3=\frac{2\pi^2}{45}\,106.75$, which amounts to about $3.2\cdot10^{-18}$.
 The next factor is the cosmological time in units of the electroweak temperature, which is  long and about \be T_{EW} t_{EW}\approx 2.2\cdot10^{15} \ee
 The fourth (last) bracket is the available time till freezeout normalized to the total time. 
 Using Friedmann evolution numbers in Appendix A one gets
   \be   { t_{FO}- t_{EW} \over t_{EW}}\approx 0.5 \ee 
 
 Since the entropy in the adiabatic expansion of the Universe is conserved, it is the same at the BBN time
 which is mostly in form of black body photons. Standard  Bose gas relation between the entropy density  and the  
 photon density is  $n_\gamma=0.1388 s_\gamma$. Substituting all these estimates in (\ref{ENT}) gives
 the baryon-to-photon ratio
 \be
  \bigg({n_B \over n_\gamma}\bigg) =7.6\cdot 10^{-2} A_{CP}
 \ee
 Since the phenomenological value for this ratio, from the BBN fits, is
 known to be
 \be 
 \bigg({n_B \over n_\gamma}\bigg)_{BBN}= 6\cdot 10^{-10}
 \ee
we conclude that the  amount of  CP violation required  to produce the observed BAU is
     
   \be   
   \label{ACPX}
   A_{CP}\approx 0.8\cdot 10^{-8} 
   \ee
 Our estimates of the CP asymmetry above gave about $A_{CP}\sim  10^{-9}$,
 an order of magnitude smaller than needed to explain BAU.
We think that this discrepancy is still inside the uncertainty of our (quite crude) estimates (\ref{ACPX}).

\section{Helical magnetogenesis} 
\label{helical}

The symmetry breaking by the Higgs VEV at $T<T_c$ leads to mass separation of the 
original non-Abelian field $A_\mu^3$ into a massive $Z_\mu$ and a massless $a_\mu$,
related by a rotation involving the Weinberg angle. The expanding outer shell of the sphaleron explosion
contains massless photons and near-massless quarks and leptons $u,d,e,\nu$.

The anomaly relation 
implies that the non-Abelian Chern-Simons number during the explosion defines the  
chiralities of the light fermions, which can be transferred  to the so called magnetic helicity

\be 
\label{HEL}
\int d^3 x A B\sim B^2 \xi^4 \sim {\rm const }
\ee  
The configurations with nonzero (\ref{HEL}) are called  {\em  helical}. We conclude that 
the primordial sphaleron explosions may {\it seed} the helical clouds of primordial magnetic fields.
Since the sphaleron rate is small, $\Gamma/T^4 < 10^{-7}$, these seeds are 
produced independently from each other, as spherical shells expanding  luminally.

\subsection{The ``inverse cascade" of magnetic fields} 
\label{magnetic_cascade}


The requirement for the inverse cascade effect is chiral unbalance which is at the origin of the CME.
Locally the trapped and co-moving light fermions produced by the sphaleron explosion are {\it chiral}. 
The time during which chirality is conserved is given by the appropriate fermion masses. For magnetic 
fields it is the electron mass, which at the sphaleron freezeout time is 

\be 
{ m_e(T_{FO})}=m_e  {v(T_{FO}) \over v(0) }  \sim  20\, {\rm KeV} 
\ee

The size growth of the chiral (linked) magnetic cloud is diffusive. For a magnetically driven plasma with a
large electric conductivity $\sigma$, a typical magnetic field $\vec B$ diffuses as

\be
\frac {d\vec B}{dt}=D\nabla^2 \vec B
\ee
with the diffusion constant $D=1/(4\pi \sigma)\sim 1/T$. It follows that the magnetic field size grows as 

\be 
R^2(t) =D \Delta t \sim {\Delta t \over T}
\ee 
where the inverse cascade time  $\Delta t  $ is  limited by the electron mass

\be 
\Delta t \sim 1/ m_e(T_{FO}) 
\ee
As a result,  the size of the chiral magnetic cloud is 
 
 \be
 R(\Delta t)\sim \bigg({1 \over m_e(T_{FO}) T}\bigg)^{\frac 12} \sim 4\cdot  {\rm fm }
 \ee 
 We note that this is  few orders of magnitude larger than  the UV scale of the problem  $1/T\sim 0.001\, {\rm fm}$, 
 and far from the  IR cutoff of the problem, the horizon at  $\sim 2.7\,{\rm  mm}$.

\subsection{CP violation results in the helical asymmetry of magnetic clouds} 
\label{CP_helical}
 
 One of the chief observation in section~\ref{CP_scale_dependence} is that 
  the magnitude of CKM induced CP violation is strongly scale dependent. It
  increases with the sphaleron size to a maximum as large as 
  ${\rm max}\,P_{CP}\sim 10^{-6}$.  Therefore, the  sphaleron seeded magnetic clouds 
  would start with such an initial asymmetry. Their subsequent evolution goes beyond 
  the scope of this work. However, we expect that during the evolution  the left- and 
  right-linked clouds to annihilate. Since helicity in magneto-hydrodynamics is conserved,
  we expect the asymmetry to grow with time.

 After the CME is switched off, ordinary magneto-hydrodynamical evolution
continues to expand the cloud size and to decrease its field  strength. This 
evolution is stopped only when the matter is no longer a plasma, 
 that is at the recombination era.



\section{Summary} \label{summary}

The main purpose of this paper is to revive  the discussion of  the cosmological EWPT, in connection to generation of the baryon asymmetry and helical magnetic clouds. 
In contrasrt to many other works, we have kept our analysis within the minimal SM,  using the 
established fact , from lattice  simulations, that the transition is a smooth cross-over.  The Higgs VEV in it is  gradually growing,  instead
of abruptly jumping, as in the previous first order scenarios.

We have focused on the primordial dynamics of the sphaleron explosions. By now, their overall  rate is more or 
less understood,  both in the symmetric and slightly broken phases, from lattice simulations. We have 
used this knowledge to study the sphaleron size distribution, by constraining the small and large 
$\rho$-tail distribution  to known results.
 
 The small-size end of the sphaleron size distribution, at $\rho \sim 1/(40\, {\rm GeV})$ was found to dominate
 the production of sound waves, as well as direct gravitational radiation. These sound waves
 may or may not be involved in the inverse acoustic cascade, advocated in  
 \cite{Kalaydzhyan:2014wca}. However if they do, long wave-length sounds would reach the horizon at the time, and
 then be converted to gravity waves, in a frequency range accessible by eLISA.
 
  In a specific time range between the transition and sphaleron freezeout $t\in [t_{EW},t_{FO}]$,
 we showed that  all three Sakharov conditions are satisfied, so the Standard Model does  generate $some$ baryon asymmetry. The magnitude depends crucially on the CP violation
 during the sphaleron explosion process.
 
 We started our studies of CP violation from
 a straightforward estimates of diagrams containing four $W$ (and CKM matrices) with
 two more $Z$ bozon added to get a nonzero result. We used   quark Dirac eigenstates
 as a generic basis. The results should still be convoluted with a spectral density for the 
 particular background. Because of the various interactions with ambient gluons, 
   the quark Dirac eigenspectrum 
 is mostly located at $\lambda \sim  M_{KW}$ (the effective mass generated by the forward scattering off gluons). If so, 
  the resulting CP asymmetry is about 10 orders of magnitude smaller than 
 needed   for the observed BAU ratio. This is a well known problem, resulting in  pessimistic view of the 
 whole approach.
 
 However, the so called ``topological stability" comes to the rescue. There are good reasons to believe that  the Dirac operator  in the background of a sphaleron explosion still possesses
a  topological zero mode,  surviving gluon rescattering. This in turn implies that the only place where
 the Klimov-Weldon mass appears is in the effective mass term for left-handed quarks, as$M_q^2/M_{KW}$.  
 These (flavor-dependent) mass contributions  cause
 additional phase shifts in the outgoing quark waves during their  production process. 
Moderatly  involved calculations of the resulting CP asymmetry set 
 its value at  about $\sim  10^{-9}$, suppressed by the Jarlskog 
 combination of CKM phases and the fourth powers of the corresponding quark masses.
  
 Comparing to what is needed to solve  
 famed BAU problem,  it is about an order of magnitude off. 
We think it is well inside the uncertainties of
  our crude estimates. Anyway, we have shown that  
 minimal standard model can generate BAU 
   many orders of magnitude larger than previously expected. 
 Clearly, further scrutiny  of this scenario is needed. 
 
 Finally, we have shown
 that like the BAU, CP asymmetry at sphaleron explosions should also be  the origin of   helical magnetic fields. The conservation of the (Abelian version) of Chern-Simons number,
 magnetic linkage, should then keep it till today, and so potentially observable.

\vskip 1cm

{\bf  Acknowledgements.\,\,} We are grateful to M.~Shaposhnikov 
who patiently criticized earlier versions of this paper. 
The work is  supported  by the U.S. Department of Energy, Office of Science, under Contract No. DE-FG-88ER40388.

\appendix

\section{Basics of Electroweak phase transition} \label{appendix_EWPT}

The transition temperature for the electroweak symmetry breaking
was known from the mean field analysis of the Higgs potential, and was further detailed 
by lattice studies in~\cite{DOnofrio:2014rug}. It is a crossover transition at 

\be 
T_{EW}=(159 \pm 1) \, {\rm GeV }
\ee
The temperature of the Universe today is $T_{\rm now}=2.73 K$. The ensuing redshift 
z-factor is

\be 
z_{EW}={T_{EW}\over T_{\rm now}}\approx 6.8\cdot10^{14} 
\ee
During the radiation dominated era, the  relation of time to temperature  is given by Friedmann
relation

\be 
t=\bigg({90 \over 32\pi^3 N_{\rm DOF}(t)}\bigg)^{\frac 12}{M_P \over T^2} 
\ee
Inserting the Planck Mass $M_P=1.2\cdot 10^{19}\, {\rm GeV}$, the transition temperature and the effective number
of degrees of freedom $N_{\rm DOF}=106.75$, we find  the time after the Big Bang to be

\be t_{EW}\sim 0.9\cdot 10^{-11} s, \,\,\,\,
c t_{EW}\approx 2.7\,{\rm  mm} \ee
As explained in the main text, the main phenomena discussed happen near the 
``sphaleron freezeout" time, which, according to Ref\cite{DOnofrio:2014rug}, is at
$T_{FO}\approx 130\, {\rm GeV}$. The corresponding cosmological time is then 
\be t_{FO}\sim 1.36\cdot 10^{-11} s, \,\,\,\,
c t_{FO}\approx 4\cdot \,{\rm  mm} \ee

The Higgs VEV $v(T)$ grows gradually, from zero at the critical $T_{EW}$. 
It was confirmed by \cite{DOnofrio:2014rug} that the squared Higgs VEV    grows approximately linearly
\be 
{v^2(140\, {\rm GeV}< T<T_{EW})\over T^2}\approx 9\bigg(1-{T\over T_{EW}}\bigg) 
\label{lattice_v_of_T} 
\ee 
This scaling is consistent with the 
naive Landau-Ginzburg treatment of the Higgs potential. The coefficient is also in agreement
with the two-loop perturbative calculations. At freezeout its value is \be v(T_{FO})\approx 167\, {\rm GeV} \ee approximately 2/3 of the value in the fully broken phase. 

In the symmetric phase $T>T_{EW}$,  the  normalized sphaleron rate remains constant, which according to \cite{DOnofrio:2014rug} is
\be {\Gamma \over T^4} \approx 1.5\cdot 10^{-7} \ee
consistent with the expected magnitude of $18\alpha_{EW}^5$ from perturbative calculations.   


If the seeded magnetic field would be simply produced at the electroweak scale $ T_{EW}$,
and then just grow with the Universe with the redshift factor $ z_{EW}$,
its  resulting spatial scale today would be

\be 
\xi\sim {z_{EW} \over T_{EW}}= 6.8\times 10^{14}\times 10^{-18}\,{\rm m} \approx 0.7\,{\rm  mm }
\ee 

The primary phase of the inverse magnetic cascade can only reach from
the micro scale of $1/ T_{EW}\sim 0.001 \,{\rm fm}$ to the horizon at that time, $c  t_{EW}$, 
about 13 orders of magnitude away. If that would be the end of the inverse cascade, the correlation length
of the magnetic chirality would be 

\be 
\xi\sim {z_{EW} \over {1/c t_{EW}}}\sim  6.8\cdot 10^{14}\times 2.7\cdot 10^{-4}\,{\rm m}\approx 10^{12}\,{\rm m }
\ee
This distance may appear large on a human scale, but in units used for intergalactic distances 
it is tiny $\frac 13 \times 10^{-11} \,{\rm Mpc}$. This scale is also the same as the predicted  
maximal wavelength of the gravity waves emitted at electroweak transition today, in 
the hypothetical inverse acoustic cascade \cite{Kalaydzhyan:2014wca}.

\section{Pure gauge sphalerons and their explosion} \label{appendix_pure_gauge_sphalerons}

Both static and time-dependent exploding solutions for the pure-gauge sphaleron have been originally discussed 
by Carter, Ostrovsky and Shuryak (COS) ~\cite{Ostrovsky:2002cg}. Its simpler derivation, to be used below, 
has been discussed by Shuryak and Zahed~\cite{Shuryak:2002qz}.
The construction relies on an  {\em off-center conformal transformation} of the $O(4)$ symmetric Euclidean instanton
solution, which is analytically continued to  Minkowski space-time. The focus of the work in
\cite{Shuryak:2002qz} was primarily  the detailed description of the fermion production. 

The original $O(4)$-symmetric solution is given by the following ansatz

\begin{eqnarray}
&&g A_\mu^a=\eta_{a\mu\nu} \partial_\nu F(y)\nonumber\\
&& F(y)=2\int_0^{\xi(y)} d\xi'   f(\xi')     
\end{eqnarray}
with $\xi=  {\rm Log}(y^2/\rho^2)$ and $\eta_{a\mu\nu}$ the 't Hooft symbol. 
Upon substitution of the gauge fields in  the gauge Lagrangian 
one finds  the effective  action for $f(\xi)$ 

\be 
S_{\rm eff}=   \int d\xi \left[{\dot{f}^2\over 2}+2f^2(1-f)^2 \right]
\ee   
corresponding to the motion of a particle in a double-well potential. 
In the Euclidean formulation, as written, the effective potential is inverted

\be 
V_E=-2f^2(1-f)^2 
\ee
and the corresponding solution  is the well known BPST instanton, 
a path connecting the two maxima of $V_E$, at $f=0,1$. 
Any other solution of the  equation of motion
following from $S_{\rm eff}$
obviously generalizes to a solution of the Yang-Mills equations for $A_\mu^a(x)$ as well.
The sphaleron itself is the static solution at the top of the potential between the minima with $f=-1/2$.

The next step is to perform an off-center conformal transformation 

\be 
(x+a)_\mu={2 \rho^2 \over (y+a)^2} (y+a)_\mu
\ee
with $a_\mu=(0,0, 0, \rho) $.  It changes the original spherically symmetric 
solution 
to a solution of the Yang-Mills equation depending on the new coordinates
 $x_\mu$, with separate dependences on time
$x_4$ and the 3-dimensional radius $r=\sqrt{x_1^2+x_2^2+x_3^2}$. 

The last step  is the analytic continuation to Minkowski time $t$, via $x_4\rightarrow i t$. 
The original parameter $\xi$ in terms of 
these   Minkowskian coordinates, which we still call  $x_\mu$, has the form

\be 
\xi ={1\over 2}  {\rm Log}\bigg({y^2\over \rho^2}\bigg)={1\over 2} {\rm Log}\left( {(t+i\rho) ^2-r^2 \over (t-i\rho) ^2-r^2 } \right)
 \ee
which is pure imaginary.To avoid carrying the extra $i$, we use the real substitution

\be 
\xi_E \rightarrow -i \xi_M ={\rm  arctan}\left( { 2 \rho t \over t^2-r^2-\rho^2 } \right)   
\label{arctan}  
\ee 
and in what follows we will drop the suffix $E$.
Switching from imaginary to real $\xi$,  correponds to switching from the Euclidean to
 Minkowski spacetime solution. It changes the sign of the acceleration, or the sign of the effective potential $V_M=-V_E$,
to that of the normal double-well problem.

The needed solution of the equation of motion has been given in ~\cite{Shuryak:2002qz} 
\footnote{There was a misprint in the index of this expression in the original paper.} 

\begin{widetext}
\be 
f(\xi)={1 \over 2} \left[ 1- \sqrt{1+\sqrt{2\epsilon}} \,{\rm dn} \left(  \sqrt{1+\sqrt{2\epsilon}} (\xi-K), {1 \over \sqrt{m}} \right) \right] 
\ee
\end{widetext}
where  ${\rm dn}(z,k) $ is one of the elliptic Jacobi functions, $2\epsilon=E/E_s,2m=1+1/\sqrt{2\epsilon}$,
and $E=V(f_{in})$ is the conserved energy of the mechanical system normalized to that of the sphaleron 
energy $E_s=V(f=1/2)=1/8$. Since the start from exactly the maximum takes a divergent time, we will start  
by {\it pushing} the sphaleron from nearby the turning point  with

 \be 
 f(0)=f_{\rm in}={1\over 2} - \kappa, \,\,\,\,\,\, f'(0)=0 
 \ee
The small displacement $\kappa$ ensures that ``rolling downhill" from the maximum takes a finite time and
that the half-period $K$ -- given by an elliptic integral -- in the expression is not divergent. 
In the plots below we will use $\kappa=0.01$, but the results  dependent on its value very weakly.

\begin{figure}[h]
\begin{center}
\includegraphics[width=7.cm]{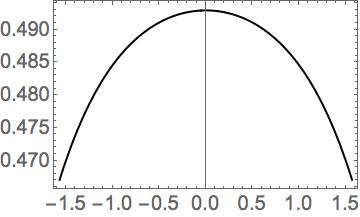}
\caption{The function $f(\xi)$ in the needed  range of its argument $\xi \in [-\pi/2,\pi/2]$}
\label{fig_f[ksi]}
\end{center}
\end{figure}

The solution above describes a particle tumbling periodically between two turning points, 
and so the expression above defines a periodic function for all $\xi$. However, as 
it is clear from (\ref{arctan}), for our particular application the only relevant domain is $\xi \in [-\pi/2,\pi/2]$. The solution $ f(\xi)$ in it 
is shown in Fig.~\ref{fig_f[ksi]}. Using the first 3 nonzero terms of its Taylor expansion

\begin{eqnarray}
f \approx &&0.49292875 - 0.0070691232 \xi^2 \nonumber\\
&&- 0.0011773\xi^4   - 0.0000781531899\xi^6 \nonumber\\
\end{eqnarray}
we find a parametrization with an accuracy of $10^{-5}$, obviously invisible in the plot and
 more than enough for our considerations. 
 
The components of the gauge potentials have the form~\cite{Shuryak:2002qz}

\begin{widetext}
\begin{eqnarray} 
&&gA_4^a=-f(\xi) { 8 t\rho x_a \over [(t-i\rho)^2-r^2]  [(t+i\rho)^2-r^2]  } \label{eqn_field} \nonumber\\
&&gA^a_i=4\rho f(\xi) { \delta_{ai}(t^2-r^2+\rho^2)+2\rho \epsilon_{aij} x_j +2 x_i x_a \over [(t-i\rho)^2-r^2]  [(t+i\rho)^2-r^2]  } 
\end{eqnarray}
which are  manifestly real.  
From those potentials we have
 generated rather lengthy expressions for the electric and magnetic fields,  and eventually for
 the CP-violating operators using Mathematica. 

Let us only mention that for the sphaleron solution itself at $t=0$, the static solution is purely magnetic with  $gA_4^a=0$.
The magnetic field squared is spherically symmetric and simple

\be 
\vec{B}^2={96 \rho^4 \over (\rho^2+r^2)^4 } 
\ee
We note that the specific expressions for pure-gauge sphaleron explosions 
  were compared with numerical real-time simulations \cite{cold_scenario_simulations} where they occur
  inside  the ``hot spots" with very good agreement~\cite{Flambaum:2010fp}.
 In the ``cold scenario" numerically studied the sphaleron size was not determined by the Higgs VEV in the broken phase, but by the size of the
  hot spots with the unbroken phase. Unfortunately, a large size tail of the sphaleron distribution on which we focused in this work 
cannot be studied in similar simulations, as their probability is prohibitively low to reach it statistically.

\section{Sphalerons dominated by Higgs VEV} 
At $T$ somewhat below $T_{EW}$, when the Higgs VEV $v(T)$ is sufficiently developed,
one may return to the original expressions developed by Klinkhamer and Manton
\cite{Klinkhamer:1984di}, modified from $T=0$ by using appropriate renormalized parameters. 
With two profile functions, $f(\xi),h(\xi)$ of normalized distance $\xi=g v r$, the sphaleron mass is  given by the following integral
\be M={4\pi v \over g} \int_0^\infty d\xi \big[ 4 (f')^2+{8\over \xi^2} f^2 (1-f)^2+{\xi^2 \over 2} (h')^2+h^2 (1-f)^2+{\xi^2\over 4} {\lambda \over g^2}(h^2-1)^2 \big]
\label{eqn_mass}\ee

The mass and size scales include temperature-dependent $v(T)$ which we took from
the lattice  simulation \cite{DOnofrio:2014rug}
\be {v(T)\over T}\approx 3 \sqrt{1-{T \over T_{EW}}} \ee 
Renormalization of all Standard Model parameters 
at finite temperatures near  $T_{EW}$ has been evaluated, via dimensional reduction, in the fundamental paper \cite{Kajantie:1995dw}. From it, extrapolated to physical Higgs mass in vacuum, we extracted, at $T$ of interest, 
the following values of the coupling
\be {\bar \lambda_3 \over \bar g_3^2} \approx 0.22,  \,\,\,\,  \,\,\,\, \bar g_3^2\approx 0.39 \ee
and ignore their  running in the temperature interval of interest. 

For calculation we use the so called ansatz B of \cite{Klinkhamer:1984di} with a single parameter $R$ 
\be f(\xi)={\xi^2 \over R(R+4)} , \,\,\, h(\xi)={\sigma R+1 \over \sigma R+2} {\xi \over R}, \,\,\, (\xi<R) \ee
\be  f(\xi)= 1-{4 \over R+4} e^{(R-\xi)/2} , \,\,\, h(\xi)=1-{R \over \sigma R +2 }{ 1\over \xi} e^{\sigma(R-\xi)} , \,\,\, (\xi>R) \ee
with $\sigma = \sqrt{2\cdot \lambda/g^2}$. These functions are plotted in Fig.\ref{fig_f_h}(a),
which, among other features, show their continuity at $\xi =R$. Putting these profiles into  the functional
(\ref{eqn_mass}), one obtains 
the sphaleron mass $M(R)$. We also calculated r.m.s. radius of the sphaleron $\rho(R)$, defined by inserting extra
$\xi^2$ into the energy density. In the main text we use $M(\rho)$  representation, with $R$ as a parameter.   

\begin{figure}[htbp]
\begin{center}
\includegraphics[width=6cm]{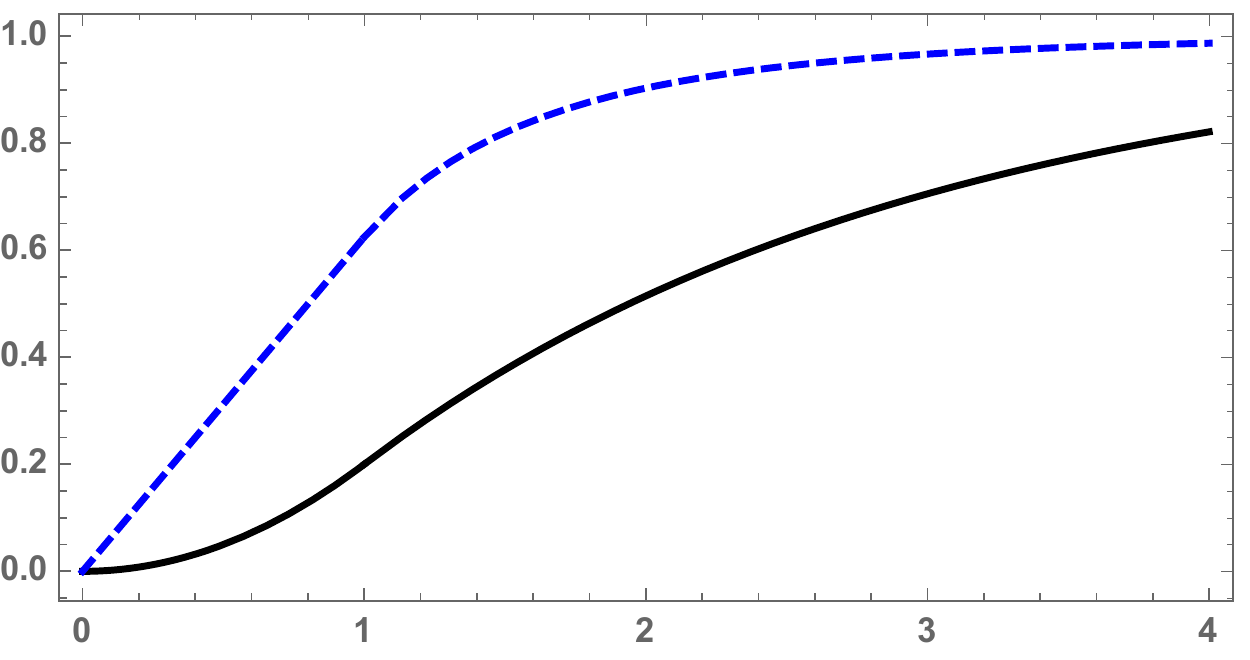} 
\includegraphics[width=6cm]{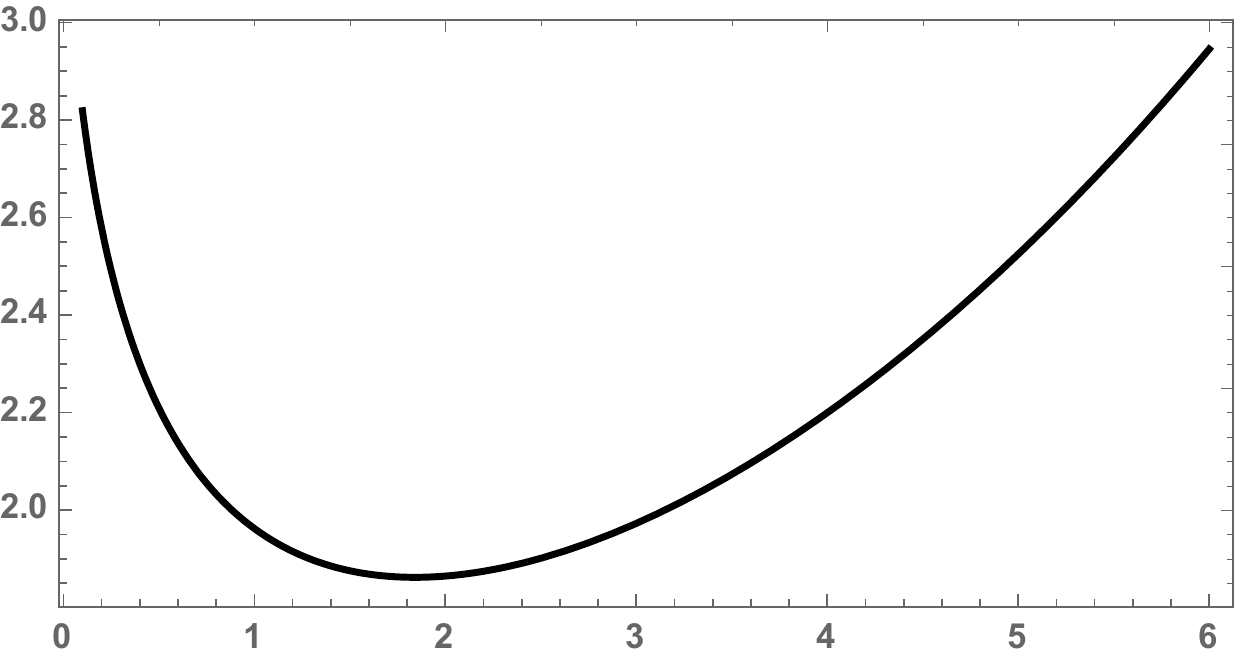} 
\caption{(a) The profile functions $f(\xi), h(\xi)$ versus $\xi$, for $R=1$, shown by black solid and blue dashed lines, respectively. (b) Root-mean-square size $\rho(R)$ as a function of parameter $R$.  }
\label{fig_f_h}
\end{center}
\end{figure}

\section{CP violation and differences of quark phases during quark production}
The multiplication of four CKM matrices by propagators, containing additional phases induced
by the quark mass terms in the Dirac operator, lead to the following expressions
$$ \Delta P_t=2  J e^{-i \phi_c-i\phi_t-i\phi_u}\big ( e^{i\phi_d}- e^{i\phi_s}\big)  ( e^{i\phi_c}- e^{i\phi_t}\big)  ( e^{i\phi_c}- e^{i\phi_u}\big)  ( e^{i\phi_t}- e^{i\phi_u}\big) $$
 $$ 
 \Delta P_c=2  J e^{-i \phi_c-i\phi_t-i\phi_u}\big ( e^{i\phi_b}- e^{i\phi_d}\big)  ( e^{i\phi_c}- e^{i\phi_t}\big)  ( e^{i\phi_c}- e^{i\phi_u}\big)  ( e^{i\phi_t}- e^{i\phi_u}\big) $$
 
$$ \Delta P_u=2  J e^{-i \phi_c-i\phi_t-i\phi_u}\big ( e^{i\phi_b}- e^{i\phi_s}\big)  ( e^{i\phi_c}- e^{i\phi_t}\big)  ( e^{i\phi_c}- e^{i\phi_u}\big)  ( e^{i\phi_t}- e^{i\phi_u}\big) $$

 $$\Delta P_b=2 J e^{-i \phi_c-i\phi_t-i\phi_u}\big ( e^{i\phi_d}- e^{i\phi_s}\big) 
  ( e^{i\phi_c}- e^{i\phi_t}\big)  ( e^{i\phi_c}- e^{i\phi_u}\big)  ( e^{i\phi_t}- e^{i\phi_u}\big) $$
  
 $$ \Delta P_s=2 J e^{-i \phi_c-i\phi_t-i\phi_u}\big ( e^{i\phi_b}- e^{i\phi_d}\big)
   ( e^{i\phi_c}- e^{i\phi_t}\big)  ( e^{i\phi_c}- e^{i\phi_u}\big)  ( e^{i\phi_t}- e^{i\phi_u}\big) $$
   
  $$ \Delta P_d=2 J e^{-i \phi_c-i\phi_t-i\phi_u}\big ( e^{i\phi_b}- e^{i\phi_s}\big)
    ( e^{i\phi_t} -e^{i\phi_c}\big)  ( e^{i\phi_c}- e^{i\phi_u}\big)  ( e^{i\phi_t}- e^{i\phi_u}\big) $$
 with
 \be J =  cos(\theta_{12}) cos(\theta_{13})^2 cos(\theta_{23}) sin(\theta_{12}) sin(\theta_{13}) sin(\theta_{23}) sin(\delta) \ee
 \end{widetext}
The squared cos is not a misprint. 
 The structure of these expressions is a reminder of the requirement that
CP violation would vanish if any pair of masses is degenerate. Indeed, in this case 
we would be able to  redefine the CKM matrix and eliminate the complex phase.


\end{document}